\documentclass[a4paper, 11pt]{article}
\usepackage{graphicx}
\usepackage{amsmath,amssymb, amsthm}
\usepackage[margin=30mm]{geometry}
\usepackage{color}
\usepackage{bbm}
\usepackage{here}
\usepackage{comment}

\usepackage{ulem}

\usepackage{enumerate}

\theoremstyle{definition}
\newtheorem{thm}{Theorem}[section]

\newtheorem{ex}[thm]{Example}

\newcommand{\INN}[2]{\langle #1, #2 \rangle}

\begin{document}
\begin{titlepage}
\null
\begin{flushright}
May, 2024\\
YGHP-24-04
\end{flushright}

\vskip 0.5cm
\begin{center}

 {\Large \bf Construction of special Lagrangian submanifolds of\\
\vskip 0.8mm
the Taub-NUT manifold and the Atiyah-Hitchin manifold 
} 
\vskip 1.0cm
\normalsize

{\bf Masato Arai${}^{a}$\footnote{arai(at)sci.kj.yamagata-u.ac.jp}
and Kurando Baba$^b$\footnote{kurando.baba(at)rs.tus.ac.jp}
}

\vskip 0.5cm

{\it $^a$Faculty of Science, Yamagata University, Kojirakawa-machi 1-4-12, Yamagata, 
Yamagata 990-8560, Japan \\
\vskip 0.5cm
$^b$Department of Mathematics, Faculty of Science and Technology, Tokyo University of Science, Noda, Chiba, 278-8510, Japan
}
\vskip 2cm

\begin{abstract}
We construct special Lagrangian submanifolds of the Taub-NUT manifold and the Atiyah-Hitchin manifold
 by combining the generalized Legendre transform approach and the moment map technique.
The generalized Legendre transform approach provides a formulation to construct hyperk\"ahler manifolds
 and can make their Calabi-Yau structures manifest.
In this approach, the K\"ahler $2$-forms and the holomorphic volume forms can be written in terms of holomorphic coordinates,
 which are convenient to employ the moment map technique.
This technique derives the condition that a submanifold in the Calabi-Yau manifold is special Lagrangian.
For the Taub-NUT manifold and the Atiyah-Hitchin manifold,
 by the moment map technique, special Lagrangian submanifolds are obtained as a one-parameter family of the orbits 
 corresponding to Hamiltonian action with respect to their K\"ahler 2-forms.
The resultant special Lagrangian submanifolds have cohomogeneity-one symmetry.
To demonstrate that our method is useful, we recover the conditions for the special Lagrangian submanifold 
 of the Taub-NUT manifold which is invariant under the tri-holomorphic $U(1)$ symmetry.
As new applications of our method, we construct special Lagrangian submanifolds of the Taub-NUT 
 manifold and the Atiyah-Hitchin manifold
 which are invariant under the action of a Lie subgroup of $SO(3)$.
In these constructions, our conditions for being special Lagrangian are expressed by 
 ordinary differential equations (ODEs) 
 with respect to the one-parameters.
We numerically give solution curves for the ODEs which specify the special Lagrangian submanifolds
for the above cases.
\end{abstract}
\end{center}
\end{titlepage}

\tableofcontents

%
%
\section{Introduction}
In the context of Riemannian geometry, Harvey and Lawson introduced a special class of minimal submanifolds 
 called calibrated submanifolds \cite{HL} (see also \cite{HL2, HL3}).
Calibrated submanifolds have attracted attention and consequently their research have been conducted both in fields  
 of mathematics and physics.
In physics, they have been studied in the context of intersecting branes 
 (see e.g. \cite{Figueroa-OFarrill:1998kci} for a review and references).
One interesting calibrated submanifold is a special Lagrangian submanifold. 
Special Lagrangian submanifolds are defined in a Calabi-Yau manifold,
 that is, submanifolds calibrated by the real part of the holomorphic volume form
 on the Calabi-Yau manifold.
Special Lagrangian submanifolds also play an important role in physics.
For example, in string theory, it is expected that explicit construction of special Lagrangian submanifolds may yield
 precise understanding of mirror symmetry \cite{SYZ, mirror2}. 
In field theory, it is shown that it appears in moduli space of a domain wall solution, which is one type of topological solitons, 
 in the nonlinear sigma model on the cotangent bundle $T^*{\bf C}P^n$ over complex projective space 
 ${\bf C}P^n$ \cite{ENOOST}.

Since the notion of special Lagrangian submanifolds was introduced, there have been many developments concerning  
 their construction.
Special Lagrangian submanifolds are given by the conditions which are constraints for coordinates of the
 Calabi-Yau manifolds.
Then, the condition for a submanifold to be special Lagrangian
 is expressed as a (nonlinear) partial differential equation.
Such an equation is difficult to solve in general.
To overcome this difficulty, we employ a method with a Hamiltonian group action,
which is called a moment map technique introduced by Joyce \cite{joyce}.
By this technique,
 we can construct special Lagrangian submanifolds which are invariant under the Hamiltonian group action
 on the ambient Calabi-Yau manifold.
Each special Lagrangian submanifold
 is obtained as a one-parameter family of the orbits for this group action.
The resultant orbits in the submanifold are of codimension one,
 so that such special Lagrangian submanifolds are said to be cohomogeneity one.
The advantage of the moment map technique is that the condition to be special Lagrangian
 is reduced to an ordinary differential equation (ODE) with respect to the one-parameter, 
 which can be solved.

With the use of the moment map technique, Joyce constructed special Lagrangian submanifolds in 
 $\mathbf{C}^n(\simeq T^*\mathbf{R}^n)$. 
In this case, 
the resultant submanifold is cohomogeneity-one
 special Lagrangian submanifolds which are invariant
 under Lie subgroups of $U(n)$.
After Joyce's work, the moment map technique was applied to construction of special Lagrangian submanifolds 
 in the cotangent bundles over compact rank one Riemannian symmetric spaces 
 whose Calabi-Yau structure was given by Stenzel \cite{stenzel1, stenzel2}.
As examples of this class, cohomogeneity-one special Lagrangian submanifolds were studied 
 in the cotangent bundle $T^*S^n$ over the sphere $S^{n}$ \cite{anciaux, IM, HS, HM} and
in $T^*{\bf C}P^n$ over the complex projective space ${\bf C}P^{n}$ \cite{AB1}.
The construction of special Lagrangian submanifolds in
the cotangent bundle over a compact Riemannian symmetric space with general rank
by using the moment map technique has been studied in \cite{Koike}.

The manifolds such as $T^*{\bf C}P^n$ explained above are hyperk\"ahler manifolds, which are special types of Calabi-Yau manifolds.
Especially, $n=1$ case of $T^*{\bf C}P^n$ is called Eguchi-Hanson manifold \cite{Eguchi:1978gw}, which is a self-
 dual solution of the four-dimensional Euclidean Einstein equation.
This type of solutions falls into the so-called Gibbons-Hawking type \cite{Gibbons:1979xm}, whose metric is given as
\begin{eqnarray}
 ds^2=V(\vec{r}) d\vec{r}\cdot d\vec{r}+V^{-1}(d\varphi+\vec{A}\cdot d\vec{r})^2\,,\quad \vec{\nabla}\times \vec{A}=\vec{\nabla}V\,, \label{GH}
\end{eqnarray}
where $\varphi, \vec{r}$ are coordinates of a four-dimensional hyperk\"ahler manifold, $\vec{A}$ is a potential.
The Gibbons-Hawking type 
 includes the Taub-NUT manifold \cite{Taub:1950ez, Newman:1963yy} and
 the Atiyah-Hitchin manifold \cite{Atiyah:1988jp}.
Since the manifold of the Gibbons-Hawking type is also Calabi-Yau manifold, it is possible to construct their special Lagrangian submanifolds.
Indeed, by using the moment map technique,
special Lagrangian submanifolds
are constructed in the case when the Calabi-Yau manifold
is $T^*{\bf C}P^n$ with $n=1$ (as well as $n \neq 1$) \cite{AB1} and the Taub-NUT manifold \cite{Noda}.
In \cite{Noda}, special Lagrangian submanifolds invariant under tri-holomorphic $U(1)$ isometry, 
  which acts on the manifold Hamiltonianly, are constructed. 
In construction, non-holomorphic coordinates are used
 in order to describe the Calabi-Yau structure on the Taub-NUT manifold and the moment map corresponding 
 to the tri-holomorphic $U(1)$-isometry.
By combining this description and the moment map technique,
the condition \cite[Proposition 3.4]{Noda}
for the $U(1)$-invariant special Lagrangian submanifolds in the Taub-NUT manifold are derived.
On the other hand, since the Taub-NUT manifold also has 
a Lie subgroup of the non-tri-holomorphic $SO(3)$ isometry, $SO(2)$,
 which acts on the manifold Hamiltonianly.
Therefore, it is possible 
 to construct special Lagrangian submanifolds invariant under that isometry.
The Atiyah-Hitchin manifold also has the same isometry, so that special Lagrangian submanifolds invariant under 
 the $SO(2)$ isometry can be constructed.
However, it appears to be difficult to apply 
 the method in \cite{Noda} to construct the $SO(2)$-isometry invariant special Lagrangian submanifolds
 in the Taub-NUT manifold as well as in the Atiyah-Hitchin manifold.
In order to construct the special Lagrangian submanifolds invariant under the $SO(2)$-isometry by using the 
 moment map technique, a most useful way is to write the geometrical quantities such as K\"ahler potential 
 in terms of holomorphic coordinates.

For this purpose, one possible approach is to use the generalized Legendre transform approach,
which is known as an alternative formulation to construct hyperk\"ahler manifolds
\cite{Lindstrom:1983rt, Hitchin:1986ea, Karlhede:1986mg, Lindstrom:1987ks}.
An advantage of this approach is that the complex structure is manifest, meaning that geometrical quantities 
 such as the K\"ahler potential of the hyperk\"ahler manifold are expressed by holomorphic coordinates.
In this approach, a K\"ahler potential is constructed from the contour integration of one function with holomorphic coordinates. 
This integration is called the $F$-function (see (\ref{F-func})). 
The $F$-functions of the Eguchi-Hanson manifold and the Taub-NUT manifold are given in \cite{Hitchin:1986ea}.
The first study of the Atiyah-Hitchin manifold by using the generalized Legendre transform approach has been done in \cite{IR}.
Later, by the same approach, the Atiyah-Hitchin manifold has been restudied in \cite{Ionas1, Ionas2}.
However, there is a discrepancy between \cite{IR} and \cite{Ionas1, Ionas2}: 
The contour of the integration in the $F$-function chosen in \cite{Ionas1} is different from one of \cite{IR}.
A different contour may give a different form of the metric of the Atiyah-Hitchin manifold.
In \cite{Arai:2022xyc}, we have shown that the contour of the integration in the $F$-function proposed in \cite{IR} is correct.
We have proven that the contour in \cite{IR} gives a real K\"ahler potential which is consistent with its definition while
 the contour in \cite{Ionas1} gives a complex K\"ahler potential.
In \cite{Arai:2022xyc}, we have calculated the K\"ahler potential and the metric in terms of holomorphic coordinates with the contour 
 in \cite{IR}, which are not derived in \cite{IR}.
To utilize the moment map technique, we also need isometric actions of the $SO(2)$ which are Hamiltonian 
 with respect to the 
 K\"ahler $2$-form obtained from the K\"ahler potential in \cite{IR}.
They can be given based on the result of the K\"ahler metric which we obtained in \cite{Arai:2022xyc}.

In this paper, we construct special Lagrangian submanifolds of the Taub-NUT manifold and the Atiyah-Hitchin manifold
 by combining the generalized Legendre
 transform approach and the moment map technique.
In order to demonstrate that our method is useful, first we rederive cohomogeneity-one special Lagrangian 
 submanifolds of the Taub-NUT manifold invariant under the tri-holomorphic $U(1)$-isometry.
We also apply our method for construction of cohomogeneity-one special Lagrangian submanifolds of the Taub-NUT 
 manifold invariant under a Lie subgroup of the $SO(3)$-isometry, $SO(2)$.
Finally we construct a cohomogeneity-one special Lagrangian submanifold of the Atiyah-Hitchin manifold which is invariant 
 under the $SO(2)$-isometry, by using our method.
In these constructions, we explicitly derive the conditions which give special Lagrangian submanifolds.
As mentioned above, the resultant special Lagrangian submanifold
 is obtained as a one-parameter family of the orbits and the corresponding condition
 to be special Lagrangian are expressed by a certain ODE for the parameter.
We solve the ODE numerically and plot solution curves which specify the special Lagrangian submanifolds for the above cases.

The organization of this paper is as follows.
In Section \ref{Sec:GLT}, we review the generalized Legendre transform approach and provide the $F$-functions of the Taub-NUT
 manifold and the Atiyah-Hitchin manifold.
We derive corresponding K\"ahler metrics in terms of holomorphic coordinates, which are necessary for construction of
 special Lagrangian submanifolds.
The calculation of the metric of the Atiyah-Hitchin manifold is based on our paper \cite{Arai:2022xyc}.
In Section \ref{sec:SLS}, we recall the definition of special Lagrangian submanifolds in a Calabi-Yau manifold and explain 
 the relation to the calibrated submanifolds.
We also review the basics of the moment map technique
to construct
special Lagrangian submanifolds in a Calabi-Yau manifold equipped with a Hamiltonian group action.
We apply the method explained in Section \ref{sec:SLS} for construction of special Lagrangian submanifolds of the Taub-NUT 
 manifold in Section \ref{Taub-NUT_SLag}.
We construct them invariant under the tri-holomorphic $U(1)$ symmetry and
 a Lie subgroup of $SO(3)$-isometry, $SO(2)$.
In Section \ref{sec:AH_SLag}, we construct a special Lagrangian submanifold of the Atiyah-Hitchin manifold invariant 
 under the $SO(2)$-isometry.
Section \ref{sec:summary} is devoted to summary and conclusion.
 
%
%
\section{The generalized Legendre transform}\label{Sec:GLT}
\subsection{Brief review of the generalized Legendre transform}
We briefly explain the generalized Legendre transform construction of hyperk\"ahler manifold \cite{Lindstrom:1987ks}.
We start with a polynomial 
\begin{eqnarray}
 \eta^{(2j)}={\bar{z} \over \zeta^j}+{\bar{v} \over \zeta^{j-1}}+{\bar{t} \over \zeta^{j-2}}+\cdots +x
  +(-)^j(\cdots + t \zeta^{j-2}-v \zeta^{j-1}+z\zeta^j)\,, \label{eta2j}
\end{eqnarray}
where $z, t, \cdots, x$ are holomorphic coordinates and $\zeta$ is the coordinate of the Riemann 
 sphere $\mathbb{C}P^1=S^2$.
This polynomial is called an ${\cal O}(2j)$-multiplet.
Eq. (\ref{eta2j}) should obey the reality condition
\begin{eqnarray}
\eta^{(2j)}(-1/\bar{\zeta})=\overline{\eta^{(2j)}(\zeta)}\,. \label{real}
\end{eqnarray}
The K\"ahler potential for a hyperk\"ahler manifold is constructed from a function with $\eta^{(2j)}$:
\begin{eqnarray}
 F=\oint_C {d\zeta \over \zeta}G(\eta^{(2j)})\,, \label{F-func}
\end{eqnarray}
where $G$ is an arbitrary holomorphic (possibly single or multi-valued) function and the contour $C$ is chosen 
 such that the result of the integration is real.
We call \eqref{F-func} the $F$-function.
The $F$-function satisfies the following set of second order differential equations
\begin{eqnarray}
 &&F_{z\bar{z}}=-F_{v\bar{v}}=F_{t\bar{t}}=\cdots =(-)^jF_{xx}\,, \\
 &&F_{z\bar{v}}=-F_{v\bar{t}}=\cdots\,,\\
 &&F_{zt}=F_{vv}\quad \quad {\rm etc}\,,\\
 &&F_{zv}=F_{vz}\quad \quad {\rm etc}\,,
\end{eqnarray}
where 
\begin{eqnarray}
F_{z\bar{z}}\equiv {\partial F^2 \over \partial z \partial {\bar{z}}}\,,\quad {\rm etc.}
\end{eqnarray}
The K\"ahler potential can be constructed from the $F$-function
by performing a two dimensional Legendre
 transform with respect to $v$ and $\bar{v}$
 \begin{eqnarray}
  K(u,\bar{u},z,\bar{z})=F(z,\bar{z},v,\bar{v},t,\bar{t},\cdots,x)-uv-\bar{u}\bar{v}\,, \label{Kahler1}
 \end{eqnarray}
 together with the extremizing conditions
 \begin{eqnarray}
  &&{\partial F \over \partial v}=u\,, \label{Fv}\\
  &&{\partial F \over \partial t}=\cdots={\partial F \over \partial x}=0 \label{Ft}\,.
 \end{eqnarray}
These equations tell us that $v, \bar{v}, t, \bar{t},\cdots,x$ are implicit functions of $z,\bar{z},u,\bar{u}$.
Considering that fact,  differentiating (\ref{Fv}) and (\ref{Ft}) with respect to $z$ gives
\begin{eqnarray}
  F_{zb}+{\partial a \over \partial z}F_{ab}=0\,,  \label{Fzb}
\end{eqnarray}
where $a, b$ run over $v,\bar{v},t,\bar{t},\cdots,x$ and summation over repeated indices is assumed.
Eq. (\ref{Fzb}) yields
\begin{eqnarray}
{\partial a \over \partial z}=-F^{ab}F_{bz}\,, \label{az}
\end{eqnarray}
where we have used $F_{zb}=F_{bz}$ and $F^{ab}$ is the inverse matrix of $F_{ab}$.
On the other hand, differentiating (\ref{Ft}) with respect to $u$, we have
\begin{eqnarray}
{\partial a \over \partial u}=F^{av}\,. \label{au}
\end{eqnarray}
Eqs. (\ref{az}) and (\ref{au}) are used to derive the K\"ahler metric in terms of derivatives of $F$ with
 respect to the holomorphic coordinates.
Taking the derivatives of (\ref{Kahler1}) with respect to $z$ and $u$, we obtain
\begin{eqnarray}
 &\displaystyle {\partial K \over \partial z}={\partial F \over \partial z}\,, &\label{Kz} \\
 &\displaystyle {\partial K \over \partial u}=-v\,. &\label{Ku} 
\end{eqnarray}
Further taking the derivatives of (\ref{Kz}) and (\ref{Ku}) 
and using (\ref{az}) and (\ref{au}), 
 we have the K\"ahler metric as
\begin{eqnarray}
&&K_{z\bar{z}}=F_{z\bar{z}}-F_{za}F^{ab}F_{b\bar{z}}\,, \label{eqn:Kzzbar-F}\\
&&K_{z\bar{u}}=F_{za}F^{a\bar{v}}\,,\\
&&K_{u\bar{z}}=F^{va}F_{a\bar{z}}\,, \\
&&K_{u\bar{u}}=-F^{v\bar{v}}\,. \label{eqn:Kuubar-F}
\end{eqnarray}

\subsection{Calabi-Yau structure in the generalized Legendre transform}
We show that the K\"ahler metric \eqref{eqn:Kzzbar-F}-\eqref{eqn:Kuubar-F} for the four-dimensional case,
 which is our main focus in this paper,
 equips the Calabi-Yau structure.
 
First we recall the definition of an almost Calabi-Yau manifold.
It is a quadruple $(M, J, \omega, \Omega)$ such that $(M, J, \omega)$ is a K\"ahler manifold $M$ of complex
 dimension $n(\ge 2)$ with a complex structure $J$ and a k\"ahler 2-form $\omega$ and a non-vanishing 
 holomorphic $(n,0)$-form $\Omega$ on $M$.
If $\omega$ and $\Omega$ satisfies the relation
\begin{equation}
 {\omega^n \over n!}=(-1)^{n(n-1)/2}\left({i \over 2}\right)^n \Omega\wedge \bar{\Omega}\,, \label{eqn:CY}
\end{equation}
the $(M, J, \omega, \Omega)$ is called a Calabi-Yau manifold.
In the following, we consider $n=2$ case. 
In this case, $\omega$ and $\Omega$ are written in terms of holomorphic coordinates $(u,z)$ as
\begin{equation}\label{eqn:kealer_K}
\omega
=\dfrac{i}{2}\left\{
K_{u\bar{u}}du\wedge d\bar{u}
+K_{u\bar{z}}du\wedge d\bar{z}
+K_{z\bar{u}}dz\wedge d\bar{u}
+K_{z\bar{z}}dz\wedge d\bar{z}
\right\}\,,
\end{equation}
and $\Omega=du\wedge dz$. 
We then find the relation between $\omega$ and $\Omega$ as
\begin{align}
\omega^{2}
= \dfrac{1}{2}\det\left(\begin{array}{cc}
K_{u\bar{u}} & K_{u\bar{z}}  \\
K_{z\bar{u}} & K_{z\bar{z}}
\end{array}\right)du\wedge dz \wedge d\bar{u} \wedge d\bar{z}
=\dfrac{1}{2}\Omega\wedge \overline{\Omega}\,, \label{eq:CY4}
\end{align}
where we have used the Monge-Amp\`ere equation \cite{Bielawski}, which holds for any 4-dimensional hyperk\"ahler manifold,
\begin{eqnarray}\label{eqn:ma}
\det\left(\begin{array}{cc}
K_{u\bar{u}} & K_{u\bar{z}}  \\
K_{z\bar{u}} & K_{z\bar{z}}
\end{array}
\right)
=1\,.
\end{eqnarray}
\eqref{eq:CY4} is the condition for the Calabi-Yau manifold for the four-dimensional case and it is seen that the hyperk\"ahler manifold is a special case for the Calabi-Yau manifold.

\subsection{Taub-NUT manifold}
\begin{figure}[H]
\centering
\includegraphics[width=6cm]{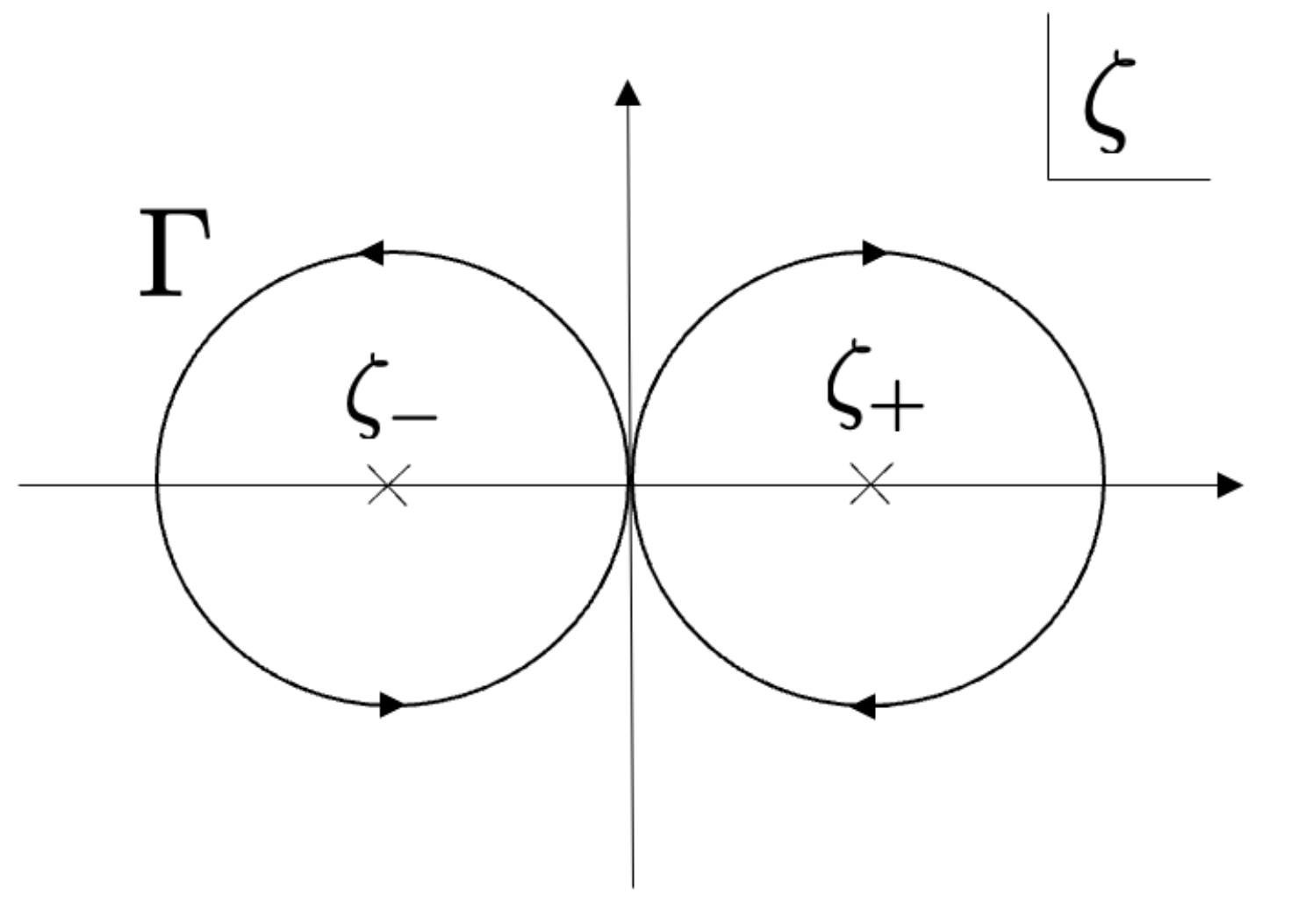}
\caption{The contour $\Gamma$.}\label{fig:TaubNUT_contour}
\end{figure}

%
%
In this subsection, we review the construction of the Taub-NUT manifold \cite{Taub:1950ez, Newman:1963yy} in the generalized Legendre transform.
First we introduce $\mathcal{O}(2)$-multiplet
\begin{align}\label{eqn:O2_multiplet}
\eta^{(2)}&=\dfrac{\bar{z}}{\zeta}+x-z\zeta=-\dfrac{z}{\zeta}(\zeta-\zeta_{+})(\zeta-\zeta_{-})\,,
\end{align}
where 
\begin{eqnarray}\label{eqn:zetapm_r}
\zeta_{\pm}=\dfrac{x\pm r}{2z}\,,\quad
r^{2}= x^{2}+4|z|^{2}\,.
\end{eqnarray}
By using this multiplet, the $F$-function of the Taub-NUT manifold is given by 
\begin{equation}\label{eqn:Ffunc_TaubNUT}
F(x,z,\bar{z})=-\dfrac{1}{2\pi i h}\oint_{\Gamma_{0}}\dfrac{d\zeta}{\zeta}(\eta^{(2)})^{2}
-\dfrac{2m}{2\pi i}\oint_{\Gamma}\dfrac{d\zeta}{\zeta}\eta^{(2)}\ln \eta^{(2)}\,,
\end{equation}
where $h$ and $m$ are constants, the latter of which is called a NUT charge.
The $\Gamma_{0}$ is a contour encircling the origin in the $\zeta$-plane counterclockwise. 
The contour $\Gamma$ is taken as in Fig. \ref{fig:TaubNUT_contour}.
We can find slightly different forms of $F$-functions for the Taub-NUT manifold
 in \cite{Hitchin:1986ea, IR}.
The difference of the $F$-functions
between \eqref{eqn:Ffunc_TaubNUT} and them
comes from the expression of the $\mathcal{O}(2)$-multiplet \eqref{eqn:O2_multiplet}.
In what follows, we will show that
the K\"ahler metric obtained from our $F$-function \eqref{eqn:Ffunc_TaubNUT}
gives a well-known form of the Taub-NUT metric.

The K\"ahler potential is obtained by the Legendre transform \eqref{Kahler1}. 
Explicitly it is given by
\begin{equation}\label{eqn:TaubNUT_Kpotantial}
K( u,\bar{u}, z,\bar{z})
=F(x, z,\bar{z})-(u+\bar{u})x\,.
\end{equation}
The extremizing condition \eqref{Fv} yields
\begin{equation}\label{eqn:Fxuub}
\dfrac{\partial F}{\partial x}=u+\bar{u}\,.
\end{equation}
Here, $(u,z)$ gives a holomorphic coordinate
on the Taub-NUT manifold.
The K\"ahler metric can be calculated from \eqref{eqn:TaubNUT_Kpotantial}:
\begin{align}
K_{z\bar{z}} &= -F_{xx}-F_{xz}F^{-1}_{xx}F_{x\bar{z}}\,,\label{eqn:Kzzb_Fxx}\\
K_{z\bar{u}} &= F_{xx}^{-1}F_{xz}\,,\label{eqn:Kzub_FxxFxz}\\
K_{u\bar{z}} &= F_{xx}^{-1}F_{x\bar{z}}\,,\label{eqn:Kuzb_FxxFxzb}\\
K_{u\bar{u}} &= -F_{xx}^{-1}\,.\label{eqn:Kzub_FuubFxx}
\end{align}
Here, we have used $F_{z\bar{z}}+F_{xx}=0$ in \eqref{eqn:Kzzb_Fxx}.
We then obtain the K\"ahler metric of the Taub-NUT manifold as
\begin{align}
K_{z\bar{z}} &= 2V+2m^{2}\dfrac{x^{2}}{r^{2}|z|^{2}}V^{-1}\,, \label{Kzzb-TN}\\
K_{z\bar{u}} &= -\dfrac{mx}{rz}V^{-1}\,,\\
K_{u\bar{z}} &= -\dfrac{mx}{r\bar{z}}V^{-1}\,,\label{Kuzb-TN}\\
K_{u\bar{u}} &= \dfrac{1}{2}V^{-1}\,, \label{Kuub-TN}
\end{align}
where
\begin{equation}
V=-{1 \over 2}F_{xx}=\dfrac{1}{h}+\dfrac{2m}{r}\,.
\end{equation}
In the following, we rewrite that the K\"ahler metric \eqref{Kzzb-TN}-\eqref{Kuub-TN}
 in terms of the spherical coordinate.
First we can show that the metric is written by the Gibbons-Hawking ansatz:
\begin{eqnarray}
 ds^2&=&K_{u\bar{u}}dud\bar{u}+K_{u\bar{z}}dud\bar{z}+K_{z\bar{u}}dzd\bar{u}+K_{z\bar{z}}dzd\bar{z} \nonumber \\
 &=&-F_{xx}dzd\bar{z}-(du-F_{xz}dz)F_{xx}^{-1}(d\bar{u}-F_{x\bar{z}}d\bar{z}) \nonumber \\
 &\sim& Vd\vec{r}\cdot d\vec{r}+V^{-1}(dp+\vec{A}\cdot d\vec{r})^{2}\,,
\end{eqnarray}
where we have used
\begin{eqnarray}
\vec{r}&=&(\tilde{x}, \tilde{y}, \tilde{z})=(z+\bar{z},-i(z-\bar{z}),x)\,,\quad |\vec{r}|=r\,, \\
p&=&{\rm Im}(u)\,,\\
\vec{A}\cdot d\vec{r}&=&{i \over 2}(F_{xz}dz-F_{x\bar{z}}d\bar{z})\,. \label{TN-metric}
\end{eqnarray}
Here the tilde symbolizes ``equal, up to an overall factor $1/2$".

We write \eqref{TN-metric} by using the spherical coordinate $(r, \theta, \phi, \psi)$:
\begin{eqnarray}
\tilde{x}&=&r\sin\theta\cos\phi\,, \\
\tilde{y}&=&r\sin\theta\sin\phi\,, \\
\tilde{z}&=&r\cos\theta=x\,, \\
\psi&=&{-p \over 2m}\,.
\end{eqnarray}
Then, we find
\begin{eqnarray}\label{metric:TN}
 ds^2=\left(1+{2m \over r}\right)(dr^2+r^2d\theta^2+r^2\sin^2\theta d\phi^2)+4m^2\left(1+{2m \over r}\right)^{-1}(d\psi+\cos\theta d\phi)^2\,.
\end{eqnarray}
where we have taken $h=1$.
This is a well-known form of the Taub-NUT metric in terms of the spherical coordinate \cite{Eguchi:1980jx}.
As mentioned in Introduction, the Taub-NUT manifold has the $U(1)$-isometry (which is tri-holomorphic) and $SO(3)$-isometry
 (which is non-tri-holomorphic). It is seen from the metric (\ref{metric:TN}).
In Section \ref{Taub-NUT_SLag}, we will construct special Lagrangian submanifolds which are invariant under $U(1)$ and a Lie
 subgroup of $SO(3)$, $SO(2)$.
%
%
\subsection{Atiyah-Hitchin manifold}\label{sec:AH_section}
In this subsection, we review the construction of the Atiyah-Hitchin manifold in the generalized Legendre transform.
This part is written based on \cite{Arai:2022xyc}.

\subsubsection{$F$-function of the Atiyah-Hitchin manifold}
\begin{figure}[H]
\centering
\includegraphics[width=7cm]{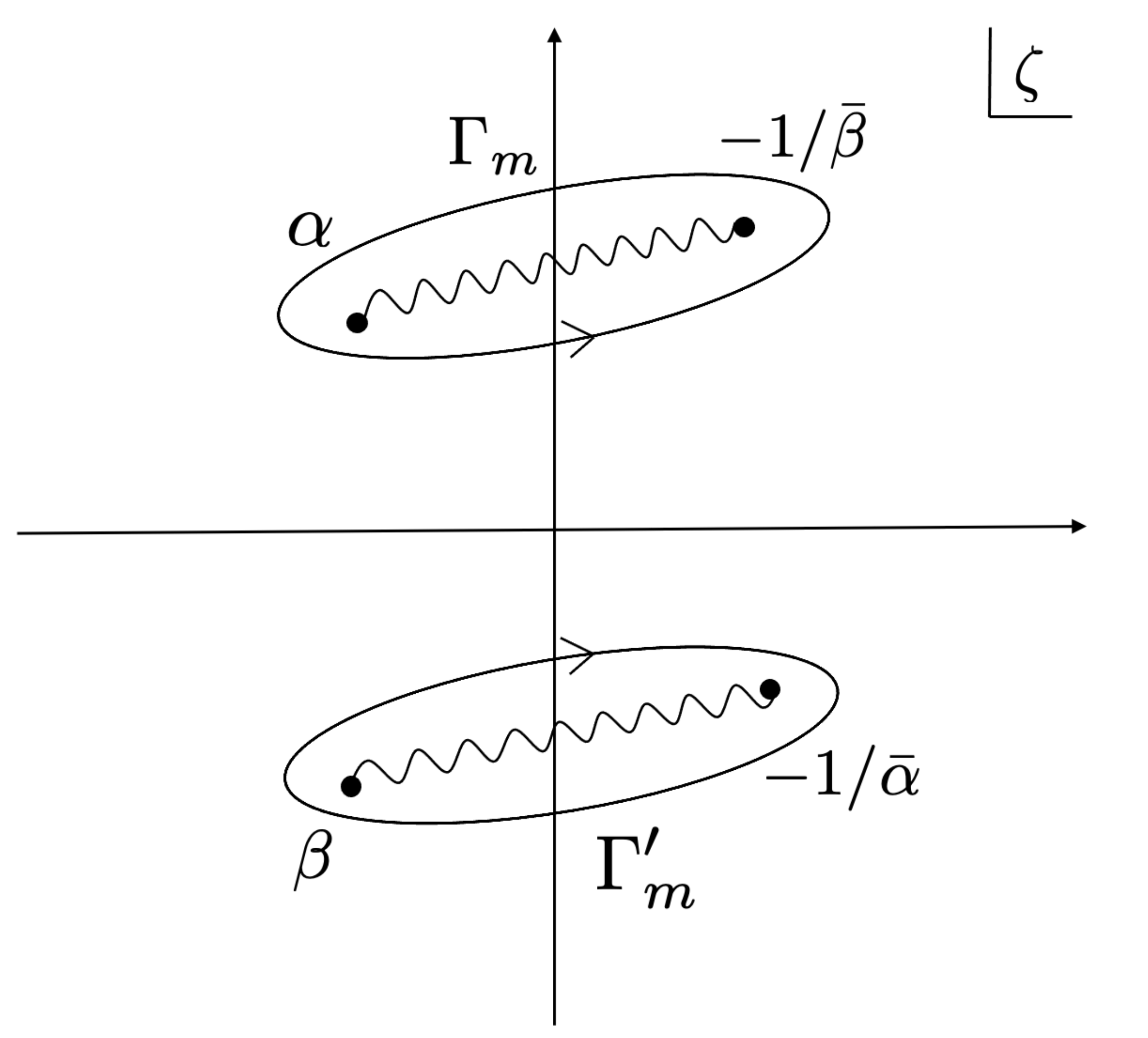}
\caption{The contour $\Gamma=\Gamma_{m}\cup\Gamma_{m}'$.}\label{fig:AH_contour}
\end{figure}

The $F$-function of the Atiyah-Hitchin manifold is written by using $\mathcal{O}(4)$-multiplet 
\begin{align}
\eta^{(4)}
&=\dfrac{\bar{z}}{\zeta^{2}}+\dfrac{\bar{v}}{\zeta}+x-v\zeta+z\zeta^{2} \notag \\
&=\dfrac{\rho}{\zeta^{2}}\dfrac{(\zeta-\alpha)(\bar{\alpha}\zeta+1)}{(1+|\alpha|^{2})}
\dfrac{(\zeta-\beta)(\bar{\beta}\zeta+1)}{(1+|\beta|^{2})}\,,
\end{align}
where $\alpha,-1/\bar{\alpha},\beta,-1/\bar{\beta}$ are the roots of $\eta^{(4)}(\zeta)=0$.
From the reality condition \eqref{real}, it is seen that $x$ is real.
The coordinates $z,v,x$ are described by $\alpha,\beta,\rho$ ($\rho>0$) as
\begin{align}
z&=\dfrac{\rho\bar{\alpha}\bar{\beta}}{(1+|\alpha|^{2})(1+|\beta|^{2})}\,,\\
v&=\dfrac{-\rho(\bar{\alpha}+\bar{\beta}-|\alpha|^{2}\bar{\beta}-\bar{\alpha}|\beta|^{2})}{(1+|\alpha|^{2})(1+|\beta|^{2})}\,,\\
x&=\dfrac{\rho(-\bar{\alpha}\beta-\alpha\bar{\beta}+(1-|\alpha|^{2})(1-|\beta|^{2}))}{(1+|\alpha|^{2})(1+|\beta|^{2})}\,.
\end{align}
The $F$-function of the Atiyah-Hitchin manifold is given by \cite{IR}
\begin{equation}
F
=F_{2}+F_{1}
=-\dfrac{1}{2\pi i h}\oint_{\Gamma_{0}}\dfrac{d\zeta}{\zeta}\eta^{(4)}
+\oint_{\Gamma}\dfrac{d\zeta}{\zeta}\sqrt{\eta^{(4)}}\,,
\end{equation}
where the contour $\Gamma_{0}$ encircles the origin of the $\zeta$-plane counterclockwise and
 the contour $\Gamma=\Gamma_{m}\cup\Gamma_{m}'$ is taken as in Fig. \ref{fig:AH_contour}.

Now we evaluate $F_1$ and $F_2$ in the $F$-function.
$F_{2}$ can be easily calculated by using the Cauchy's integral formula as
\begin{equation}
F_{2}=-\dfrac{x}{h}\,.
\end{equation}
On the other hand, a lengthy calculation is necessary for $F_1$.
In the following, we give a brief explanation of the calculation
(in more detail, see \cite{Arai:2022xyc}).

We can rewrite $F_1$ by using the integral
\begin{equation}
\mathcal{I}_{n}(\Gamma)=\oint_{\Gamma}\zeta^{n}\dfrac{d\zeta}{2\zeta\sqrt{\eta^{(4)}}}\,, \quad n=0,~1,~2\,,
\end{equation}
as
\begin{equation}\label{eqn:ABI_eq32}
F_{1}=2x\mathcal{I}_{0}(\Gamma)-2(v\mathcal{I}_{1}(\Gamma)+\overline{v\mathcal{I}_{1}(\Gamma)})+2(z\mathcal{I}_{2}(\Gamma)+\overline{z\mathcal{I}_{2}(\Gamma)})\,.
\end{equation}
\begin{figure}[H]
\centering
\includegraphics[width=10cm]{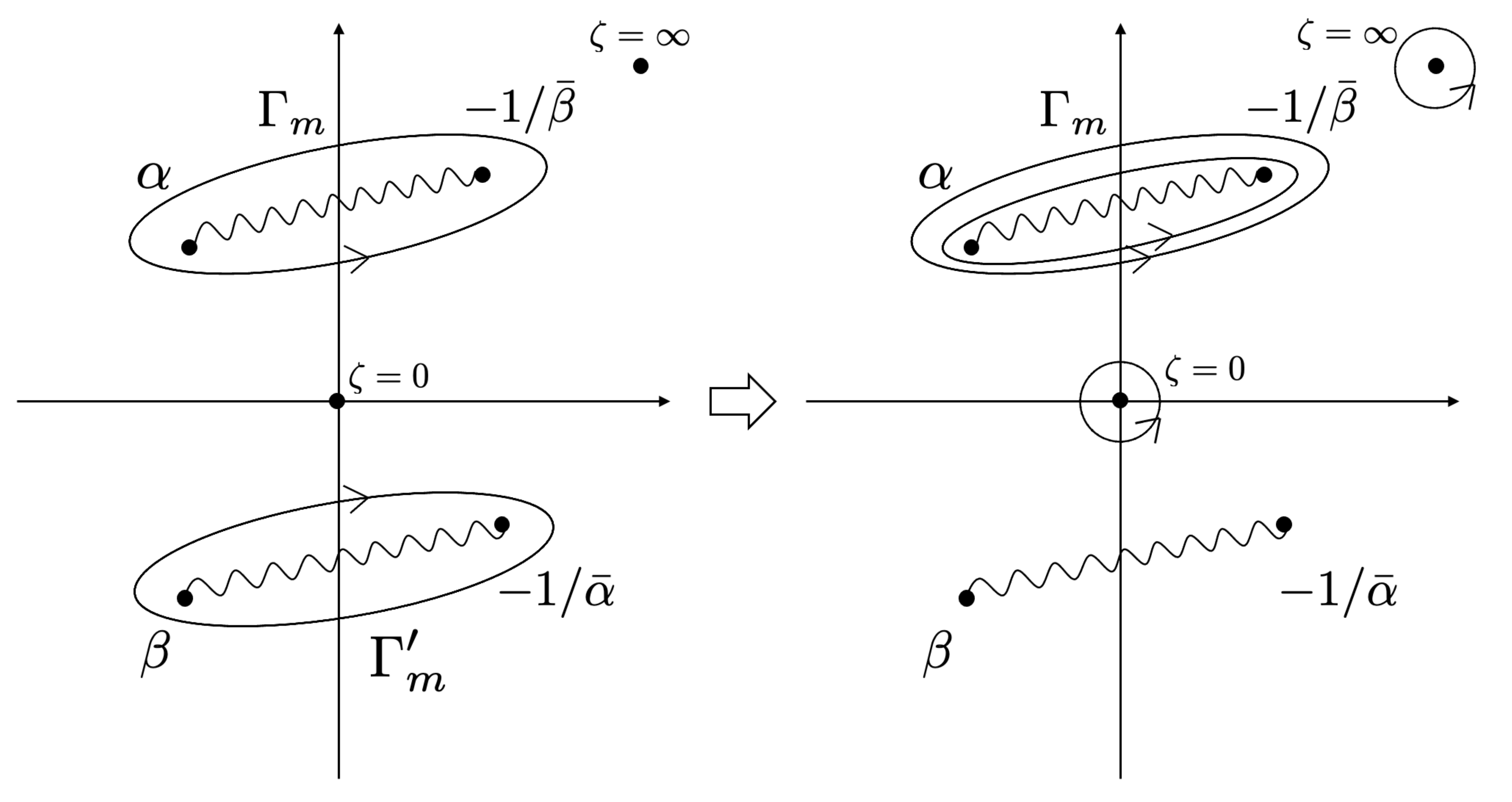}
\caption{Deformation of $\Gamma_m'$.}\label{fig:def}
\end{figure}
It is possible to deform the contour $\Gamma_{m}'$ in $\Gamma=\Gamma_{m}\cup\Gamma_{m}'$ to $\Gamma_m$ as in 
 Fig. \ref{fig:def}.
While deforming $\Gamma_m'$ to $\Gamma_m$, we need to pick up the poles of the integral $\mathcal{I}_{n}(\Gamma)$ 
 and have
\begin{equation}
\mathcal{I}_{n}(\Gamma)
=\begin{cases}
2\mathcal{I}_{0}(\Gamma_{m}) & (n=0)\,,\\
2\mathcal{I}_{1}(\Gamma_{m})+2\pi i\left(-\dfrac{1}{2\sqrt{z}}\right) & (n=1)\,,\\
2\mathcal{I}_{2}(\Gamma_{m})+2\pi i\left(-\dfrac{1}{2\sqrt{z}}\cdot\dfrac{v}{2z}\right) & (n=2)\,.
\end{cases}
\end{equation}
Substituting this into \eqref{eqn:ABI_eq32}, $F_{1}$ turns out to be
\begin{equation}\label{eqn:ABI_eq41}
F_{1}
=4\left\{
x\mathcal{I}_{0}(\Gamma_{m})-\left(
v\mathcal{I}_{1}(\Gamma_{m})-z\mathcal{I}_{2}(\Gamma_{m})-\dfrac{\pi i}{4}\cdot \dfrac{v}{\sqrt{z}}
+\text{c.c}\right)
\right\}\,.
\end{equation}
For the calculation of $F_1$, we need to evaluate $\mathcal{I}_{n}(\Gamma_{m})$ ($n=0,1,2$).
Those can be evaluated by using the elliptic function.

\begin{figure}[H]
\centering
\includegraphics[width=7cm]{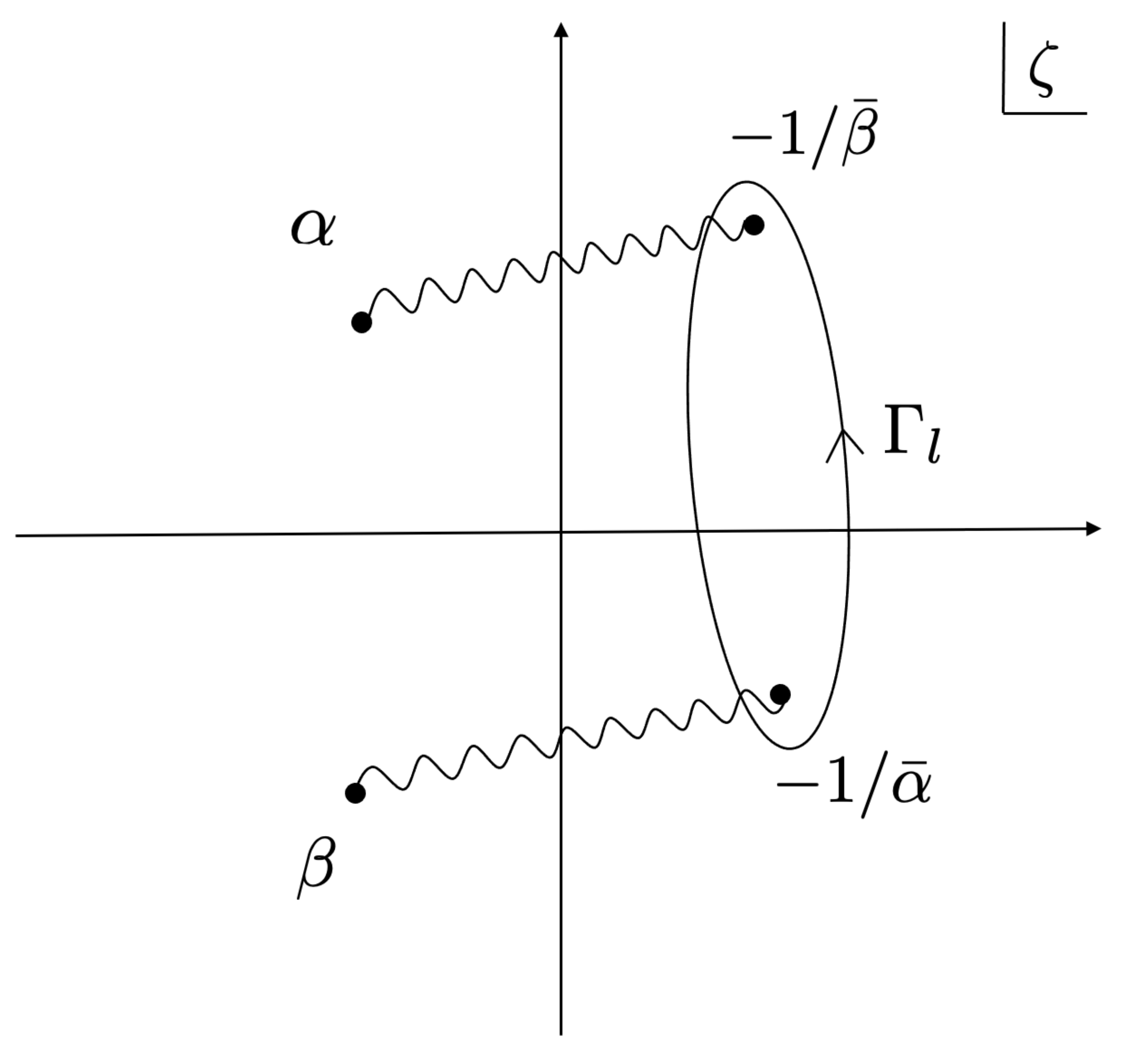}
\caption{The contour $\Gamma_{l}$.}\label{fig:ABI_fig3}
\end{figure}

To this end, we consider an elliptic curve on the $(\zeta,\eta)$ plane
\begin{equation}
C:\quad
\eta^{2}=4\zeta^{2}\eta^{(4)}(\zeta)\,.  \label{eqn:torus}
\end{equation}
This elliptic curve gives a globally defined holomorphic $1$-form $\varpi$ as follows:
\begin{equation}
\varpi=\dfrac{d\zeta}{\eta}=\dfrac{d\zeta}{2\zeta\sqrt{\eta^{(4)}}}\,.
\end{equation}
$\varpi$ is called an abel-form of $C$ and it is uniquely determined up to a constant.
Here, $\varpi$ is chosen such that it coincides with a $1$-form  to determine the integral $\mathcal{I}_{0}(\Gamma_{m})$.
The two periods of $\varpi$ are described by its integral over canonical cycles.
Those cycles are $\Gamma_{m}$ and $\Gamma_{l}$ in Fig. \ref{fig:ABI_fig3}.
The two periods $2\omega$ and $2\omega'$ are given by
\begin{equation}
2\omega=\oint_{\Gamma_{m}}\varpi\,,\quad
2\omega'=\oint_{\Gamma_{l}}\varpi\,.
\end{equation}
The half periods are written by the complete elliptic integral of the first kind $K(k)$
\begin{equation}\label{eqn:omega_1/rhoKk}
\omega=\dfrac{1}{\sqrt{\rho}}K(k)\,,\quad
\omega'=\dfrac{i}{\sqrt{\rho}}K(k')\,,
\end{equation}
where
\begin{equation}
k=\dfrac{|1+\bar{\alpha}\beta|}{\sqrt{(1+|\alpha|^{2})(1+|\beta|^{2})}}\,,
\end{equation}
and $k'=\sqrt{1-k^{2}}$.
It is seen that $\omega\in\mathbb{R}$ and $\omega'\in i\mathbb{R}$.
In the following, we express as
\begin{equation}
\omega_{1}=\omega\,,\quad
\omega_{2}=\omega+\omega'\,,\quad
\omega_{3}=\omega'\,.
\end{equation}

We rewrite the curve $C$ to the Weierstrass normal form. 
To do this, we first make use of the following birational transformation:
\begin{equation}\label{eqn:bitrans}
\begin{cases}
\nu=[\zeta,-1/\bar{\alpha},\alpha,\beta]
=\dfrac{(\zeta-\alpha)(1+\bar{\alpha}\beta)}{(\zeta-\beta)(1+|\alpha|^{2})}\,,\\
\mu=\eta\dfrac{\partial\nu}{\partial\zeta}\,.
\end{cases}
\end{equation}
Then, the curve $C$ in \eqref{eqn:torus}
is expressed as a Riemann normal form:
\begin{equation}\label{eqn:CLow}
\mu^{2}=4\rho\nu(\nu-1)(\nu-k^{2})\,,
\end{equation}
and the four roots $\alpha$, $-1/\bar{\beta}$, $-1/\bar{\alpha}$,
$\beta$ correspond to $0$, $k^{2}$, $1$, $\infty$, respectively.

Next, we change the variables $(\nu, \mu)$ to $(X,Y)$ by the following transformations:
\begin{eqnarray}
&&\nu=
\dfrac{X}{\rho}+\dfrac{1+k^{2}}{3}\,,\label{eqn:zetatoX} \\
&&\mu=\dfrac{Y}{\rho}\,.
\end{eqnarray}
The curve $C$ is then expressed as the Weierstrass normal form
\begin{align}
Y^{2} 
&=4(X-e_{1})(X-e_{2})(X-e_{3})=4X^{3}-g_{2}X-g_{3}\,,
\end{align}
where 
\begin{equation}\label{eqn:ABI_eq60}
e_{1}=-\dfrac{\rho}{3}(k^{2}-2)\,,\quad
e_{2}=\dfrac{\rho}{3}(2k^{2}-1)\,,\quad
e_{3}=-\dfrac{\rho}{3}(k^{2}+1)\,.
\end{equation}
It is verified that those roots satisfy the following relations:
\begin{equation}\label{eqn:ABI_eq63}
e_{1}+e_{2}+e_{3}=0\,,\quad
e_{1}e_{2}+e_{2}e_{3}+e_{3}e_{1}=-\dfrac{g_{2}}{4}\,,\quad
e_{1}e_{2}e_{3}=\dfrac{g_{3}}{4}\,,
\end{equation}
and
\begin{equation}\label{eqn:ABI_eq61}
e_{1}-e_{3}=\rho\,,\quad
\dfrac{e_{2}-e_{3}}{e_{1}-e_{3}}=k^{2}\,.
\end{equation}
From \eqref{eqn:ABI_eq60}, we find
\begin{equation}
g_{2}=\dfrac{4}{3}\rho^{2}(1-k^{2}+k^{4})\,,\quad
g_{3}=\dfrac{4}{27}\rho^{3}(k^{2}-2)(2k^{2}-1)(k^{2}+1)\,.
\end{equation}
Those yield the discriminant of $C$ as $\Delta=g_{2}^{3}-27g_{3}^{2}=16\rho^{6}k^{4}k'^{4}\neq 0$.
Therefore, the curve $C$ is regular.

Next we shall rewrite $\mathcal{I}_{n}(\Gamma_{m})$ to the Weierstrass normal form.
We denote by $X_{\zeta}$ the image of $\zeta$ by the transformation which combines 
 \eqref{eqn:bitrans} and \eqref{eqn:zetatoX}:
\begin{equation}\label{eqn:zeta_X}
\dfrac{(\zeta-\alpha)(1+\bar{\alpha}\beta)}{(\zeta-\beta)(1+|\alpha|^{2})}=\dfrac{X-e_{3}}{e_{1}-e_{3}}\,.
\end{equation}
Under this convention, we have
\begin{align}
X_{0} &= e_{3}+\rho\cdot\dfrac{\alpha}{\beta}\dfrac{1+\bar{\alpha}\beta}{1+|\alpha|^{2}}\,,\\
X_{\infty} &= e_{3}+\rho\cdot\dfrac{1+\bar{\alpha}\beta}{1+|\alpha|^{2}}\,.
\end{align}
Then, \eqref{eqn:zeta_X} is rewritten as
\begin{equation}\label{eqn:ABI_eq70}
\zeta=\beta\dfrac{X-X_{0}}{X-X_{\infty}}\,.
\end{equation}
By using this transformation, the abel form $\varpi$ and the integral ${\cal I}_n(\Gamma_m)$ are expressed as
\begin{eqnarray}
\varpi=\dfrac{dX}{Y}=\dfrac{dX}{\sqrt{4X^{3}-g_{2}X-g_{3}}}\,,
\end{eqnarray}
and
\begin{eqnarray}
\mathcal{I}_{n}(\Gamma_{m})&=&
\oint_{\Gamma_{m}}\left(\beta\dfrac{X-X_{0}}{X-X_{\infty}}\right)^{n}\dfrac{dX}{Y}\,, \label{eqn:ABI_eq71}
\end{eqnarray}
respectively.
Here, the contour $\Gamma_{m}$ on $\zeta$-plane is mapped to one on $X$-plane via \eqref{eqn:ABI_eq70},
 which we write the same symbol, namely, $\Gamma_{m}$.
This winds once around the branch-cut between the roots $e_{3}$ and $e_{2}$
 (resp.~the roots $e_{2}$ and $e_{1}$) on $X$-plane.

In the following, we evaluate \eqref{eqn:ABI_eq71}.
Clearly, \eqref{eqn:ABI_eq71} with $n=0$ case is the period of the torus:
\begin{equation}\label{eqn:ABI_eq72}
\mathcal{I}_{0}(\Gamma_{m})
=\oint_{\Gamma_{m}}\dfrac{dX}{Y}=2\omega_{1}\,.
\end{equation}

$\mathcal{I}_{n}(\Gamma_{m})$ with $n=1,2$ case can be evaluated by using the Weierstrass
 elliptic function.
For preparation, we briefly review the Weierstrass $\wp$-function.
Let $\Lambda$ denote the orthogonal lattice in ${\mathbb C}$ defined by
 $\Lambda=\mathbb{Z}\cdot 2\omega\oplus \mathbb{Z}\cdot 2\omega^\prime$.
The Weierstrass $\wp$-function $\wp(u)=\wp(u, \Lambda)$ is defined by
\begin{equation}
\wp(u)=\dfrac{1}{u^{2}}+\sum_{\lambda\in\Lambda-\{0\}}
\left\{
\dfrac{1}{(u-\lambda)^{2}}-\dfrac{1}{\lambda^{2}}\
\right\}\,,
\quad
u\in\mathbb{C}\,.
\end{equation}
This function is even and has the double periodicity, that is,
\begin{equation}\label{eqn:wp_dp}
\wp(u+2\omega)=\wp(u)\,,\quad
\wp(u+2\omega')=\wp(u)\,,\quad
u\in\mathbb{C}\,.
\end{equation}
$\wp(u)$ also satisfies the following differential equation:
\begin{equation}
(\wp'(u))^{2}=4\wp(u)^{3}-g_{2}\wp(u)-g_{3}\,.
\end{equation}
We denote by $C^{*}$ the projectivization of $C$, i.e.,
\begin{equation}
C^{*}
=\{[x_{0},x_{1},x_{2}]\in\mathbb{C}P^{2}
\mid
x_{0}x_{2}^{2}=4x_{1}^{3}-g_{2}x_{0}^{2}x_{1}-g_{3}x_{0}^{3}\}\,.
\end{equation}
Thanks to \eqref{eqn:wp_dp}, the $\wp$-function
 induces a function on the torus $\mathbb{C}/\Lambda$, which we write the same symbol $\wp(u)$. 
Then, we obtain a map $\psi:\mathbb{C}/\Lambda\to\mathbb{C}P^{2}$ defined by
\begin{equation}\label{eqn:abel}
\psi(u)=[1, \wp(u),\wp'(u)]=[1, X,Y]\,,\quad
u\in\mathbb{C}/\Lambda\,.
\end{equation}
$\psi$ gives an isomorphism between $\mathbb{C}/\Lambda$ and $C^{*}$.
We call this map the abel map of $C^{*}$.
Through this map, we express $u_{\zeta}\in\mathbb{C}/\Lambda$ as the element corresponding to 
 $X_\zeta$ ($\zeta\in\mathbb{C}\cup\{\infty\}$):
\begin{equation}
X_{\zeta}=\wp(u_{\zeta})\,.
\end{equation}
Furthermore, we set $Y_{\zeta}=\wp'(u_{\zeta})$.
It is convenient to divide $u_{\zeta}$ into the real part
 and the imaginary part with respect to
 the antiholomorphic involution $\zeta\mapsto -1/\bar{\zeta}$
 on $\mathbb{C}\cup\{\infty\}$, that is,
\begin{equation}
u_{\zeta}^{\pm}
=u_{\zeta}\pm u_{-1/\bar{\zeta}}\,,
\end{equation}
so that we have 
\begin{equation}\label{eqn:uinfty_pm}
u_{\infty}^{\pm}=u_{\infty}\pm u_{0}\,.
\end{equation}
We write $(x_{\pm},y_{\pm})$ as the $(X,Y)$-coordinates of the point corresponding to 
$u_{\infty}^{\pm}$ via the abel map $\psi$, \eqref{eqn:abel}. 
Then, we can prove that the following relations
\begin{equation}\label{eqn:xpm_escpt}
x_{\pm}= \dfrac{x\pm 6|z|}{3}\,,
\end{equation}
and
\begin{equation}
y_{+}=iv_{+}(x_{+}-x_{-})\,,\quad
y_{+}=v_{-}(x_{-}-x_{+})\,, \label{eqn:ypm}
\end{equation}
hold, where
\begin{equation}
v_{+}=\mathrm{Im}\dfrac{v}{\sqrt{z}}\,,\quad
v_{-}=\mathrm{Re}\dfrac{v}{\sqrt{z}}\,.
\end{equation}

Now we are ready to evaluate $\mathcal{I}_{1}(\Gamma_{m})$.
By using \eqref{eqn:ABI_eq72}, this is calculated as
\begin{align}
\mathcal{I}_{1}(\Gamma_{m})
&=\beta\left\{\oint_{\Gamma_{m}}\dfrac{dX}{Y}+\dfrac{X_{\infty}-X_{0}}{Y_{\infty}}\oint_{\Gamma_{m}}\dfrac{Y_{\infty}}{X-X_{\infty}}\dfrac{dX}{Y}\right\} \notag \\
&=2\beta\omega+\beta\dfrac{X_{\infty}-X_{0}}{Y_{\infty}}\oint_{\Gamma_{m}}\dfrac{Y_{\infty}}{X-X_{\infty}}\dfrac{dX}{Y}\,.
\label{eqn:ABI_eq83'}
\end{align}
To evaluate the second term in \eqref{eqn:ABI_eq83'}, we consider the following integral
\begin{align}
\pi(X_{\zeta})
&=-\oint_{\Gamma_{m}}\dfrac{Y_{\zeta}}{X-X_{\zeta}}\dfrac{dX}{Y}=-2\int_{e_{3}}^{e_{2}}\dfrac{Y_{\zeta}}{X-X_{\zeta}}\dfrac{dX}{Y}\,.
\end{align}
By using the abel map \eqref{eqn:abel}, this integral can be expressed by $\wp$-function as
\begin{equation}\label{eqn:ABI_85''}
\pi(X_{\zeta})
=-2\int_{\omega_{3}}^{\omega_{2}}\dfrac{\wp'(u_{\zeta})}{\wp(u)-\wp(u_{\zeta})}du\,,
\end{equation}
where we have used $e_{j}=\wp(\omega_{j})$.
With the use of the $\zeta$-function, $\zeta(u)=\zeta(u;\Lambda)$ and $\sigma$-function, $\sigma(u)=\sigma(u;\Lambda)$,
 the integrand in \eqref{eqn:ABI_85''} can be written by
\begin{align} 
\dfrac{\wp'(u_{\zeta})}{\wp(u)-\wp(u_{\zeta})}
&=-\zeta(u+u_{\zeta})+\zeta(u-u_{\zeta})+2\zeta(u_{\zeta}) \notag \\
&=-\dfrac{d}{du}\log\sigma(u+u_{\zeta})+\dfrac{d}{du}\sigma(u-u_{\zeta})+2\zeta(u_{\zeta})\,,
\end{align}
so, we find
\begin{equation}\label{eqn:ABI_eq86}
\pi(X_{\zeta})
=2\log\dfrac{\sigma(\omega_{2}+u_{\zeta})}{\sigma(\omega_{2}-u_{\zeta})}
-2\log\dfrac{\sigma(\omega_{3}+u_{\zeta})}{\sigma(\omega_{3}-u_{\zeta})}-4\omega_{1}\zeta(u_{\zeta})\,.
\end{equation}
Here, we use the monodromy property of $\sigma$-function for $j=2,3$
\begin{align}
\sigma(\omega_{j}+u_{\zeta})
=e^{2\eta_{j}u_{\zeta}}\sigma(\omega_{j}-u_{\zeta})\,,
\end{align}
so that the following relation holds:
\begin{equation}\label{eqn:logmonod}
2\log\dfrac{\sigma(\omega_{j}+u_{\zeta})}{\sigma(\omega_{j}-u_{\zeta})}
=
4\eta_{j}u_{\zeta}\quad
(\mathrm{mod}~2\pi i\mathbb{Z})\,,
\end{equation}
where $\eta_j$ is the quasi-half period of the curve $C$.
For instance, $\eta_1$ is defined by (in more detail on $\eta_j$, see Appendix A in \cite{Arai:2022xyc})
\begin{equation}
\eta_{1}=-\int_{e_{3}}^{e_{2}}X\dfrac{dX}{Y}
=-\dfrac{1}{2}\oint_{\Gamma_{m}}X\dfrac{dX}{Y}\,. \label{eq:eta_int}
\end{equation}
Substituting \eqref{eqn:logmonod} into \eqref{eqn:ABI_eq86}, we have
\begin{equation}
\pi(X_{\zeta})=4\det\left(
\begin{array}{cc}
u_{\zeta} & \omega_{1} \\
\zeta(u_{\zeta}) & \zeta(\omega_{1})
\end{array}
\right)\quad
(\mathrm{mod}~2\pi i\mathbb{Z})\,.
\end{equation}
We need to evaluate this with $\zeta\rightarrow \infty$, $\pi(X_\infty)$.
From \eqref{eqn:uinfty_pm}, we have
\begin{equation}
u_{\infty}=\dfrac{1}{2}(u_{\infty}^{+}+u_{\infty}^{-})\,,
\end{equation}
and therefore, we find
\begin{multline}\label{eqn:ABI_eq93}
\det\left(
\begin{array}{cc}
u_{\infty} & \omega_{1} \\
\zeta(u_{\infty}) & \zeta(\omega_{1})
\end{array}
\right)\\
=\dfrac{1}{2}\left\{
\det\left(
\begin{array}{cc}
u_{\infty}^{+} & \omega_{1} \\
\zeta(u_{\infty}^{+}) & \zeta(\omega_{1})
\end{array}
\right)+\det\left(
\begin{array}{cc}
u_{\infty}^{-} & \omega_{1} \\
\zeta(u_{\infty}^{-1}) & \zeta(\omega_{1})
\end{array}
\right)+\omega_{1}\dfrac{Y_{\infty}}{X_{\infty}-X_{0}}
\right\}\,.
\end{multline}
Using this equation, $\pi(X_\infty)$ can be calculated as
\begin{equation}
\pi(X_{\infty})
=\dfrac{1}{2}\left\{
\pi(x_{+})+\pi(x_{-})
\right\}+2\omega_{1}\dfrac{Y_{\infty}}{X_{\infty}-X_{0}}
\quad
(\mathrm{mod}~\pi i\mathbb{Z})\,.
\end{equation}
There exists $a\in\mathbb{Z}$ such that
\begin{equation}\label{eqn:ABI_eq95}
\pi(X_{\infty})
=\dfrac{1}{2}\left\{
\pi(x_{+})+\pi(x_{-})
\right\}+2\omega_{1}\dfrac{Y_{\infty}}{X_{\infty}-X_{0}}+a\pi i\,.
\end{equation}
We substitute \eqref{eqn:ABI_eq95} into \eqref{eqn:ABI_eq83'} and finally find
\begin{align}
\mathcal{I}_{1}(\Gamma_{m})
&=2\beta\omega_{1}-\beta\dfrac{X_{\infty}-X_{0}}{Y_{\infty}}\left[
\dfrac{1}{2}\left\{
\pi(x_{+})+\pi(x_{-})
\right\}+2\omega_{1}\dfrac{Y_{\infty}}{X_{\infty}-X_{0}}+a\pi i
\right] \notag \\
&=\dfrac{1}{4\sqrt{z}}\left\{
\pi(x_{+})+\pi(x_{-})
+2a\pi i\right\}\,,
\label{eqn:ABI_eq96}
\end{align}
where we have used the following relation in the second equality:
\begin{equation}\label{eqn:ABI_eq75_1st}
\dfrac{Y_{\infty}}{X_{0}-X_{\infty}}=2\beta\sqrt{z}\,.
\end{equation}

Next we evaluate \eqref{eqn:ABI_eq71} for $n=2$ case.
The integrand of $\mathcal{I}_{2}(\Gamma_{m})$ can be transformed as
\begin{align}\label{eqn:ABI_eq98'}
\left(\beta\dfrac{X-X_{0}}{X-X_{\infty}}\right)^{2}
=\beta^{2}&+
2\beta^{2}\dfrac{X_{\infty}-X_{0}}{Y_{\infty}}\cdot\dfrac{Y_{\infty}}{X-X_{\infty}}\notag\\
&+\beta^{2}\left(\dfrac{X_{\infty}-X_{0}}{Y_{\infty}}\right)^{2}\left(
\dfrac{Y_{\infty}}{X-X_{\infty}}\right)^2\,.
\end{align}
The integral of the first term can be easily calculated from \eqref{eqn:ABI_eq72}.
The integral of the second term is evaluated from \eqref{eqn:ABI_eq75_1st} and \eqref{eqn:ABI_eq95} as
\begin{equation}
\oint_{\Gamma_{m}}
2\beta^{2}\dfrac{X_{\infty}-X_{0}}{Y_{\infty}}\cdot\dfrac{Y_{\infty}}{X-X_{\infty}}\dfrac{dX}{Y}
=\dfrac{\beta}{2\sqrt{z}}\left\{
\pi(x_{+})+\pi(x_{-})+2a\pi i
\right\}-4\beta^{2}\omega_{1}\,. \label{eqn:int-3-2}
\end{equation}
To calculated the integral of the third term, we observe
\begin{equation}
\left(
\dfrac{Y_{\infty}}{X-X_{\infty}}\right)^{2}
=2(X-X_{\infty})-\dfrac{12X_{\infty}^{2}-g_{2}}{2Y_{\infty}}\cdot\dfrac{Y_{\infty}}{X-X_{\infty}}
-Y\dfrac{d}{X}\left(\dfrac{Y}{X-X_{\infty}}\right)\,. \label{eqn:int-def}
\end{equation}
The integral of the first term in \eqref{eqn:int-def} is found to be
\begin{equation}
\oint_{\Gamma_{m}}2(X-X_{\infty})\dfrac{dX}{Y}=-4\eta_{1}-4\omega_{1}\left(\dfrac{x}{3}-\beta v+2\beta^{2}z\right)\,, \label{eqn:def-1st}
\end{equation}
where we have used \eqref{eq:eta_int} and 
\begin{equation}
X_{\infty}=
\dfrac{x}{3}-\beta v+2\beta^{2}z\,.
\end{equation}
Using the following relation
\begin{equation}\label{eqn:12Xinfty2-g2}
\dfrac{12X_{\infty}^{2}-g_{2}}{4(X_{0}-X_{\infty})}
=\beta v-4\beta^{2}z\,,
\end{equation}
we can calculate the integral of the second term in \eqref{eqn:int-def}:
\begin{multline}
\oint_{\Gamma_{m}}
\dfrac{12X_{\infty}^{2}-g_{2}}{2Y_{\infty}}\cdot\dfrac{Y_{\infty}}{X-X_{\infty}}
\dfrac{dX}{Y}\\
=-\dfrac{1}{2}\left(\dfrac{v}{\sqrt{z}}-4\beta\sqrt{z}\right)
\left\{
\pi(x_{+})+\pi(x_{-})+2a\pi i
\right\}+4\omega_{1}(\beta v-4\beta^{2}z)\,. \label{eqn:def-2nd}
\end{multline}
The integral of the third term in  \eqref{eqn:int-def} can be evaluated as
\begin{equation}
\oint_{\Gamma_{m}}Y\dfrac{d}{dx}\left(\dfrac{Y}{X-X_{\infty}}\right)\dfrac{dX}{Y}
=\oint_{\Gamma_{m}}d\left(\dfrac{Y}{X-X_{\infty}}\right)=0\,. \label{eqn:def-3rd}
\end{equation}
From \eqref{eqn:def-1st}, \eqref{eqn:def-2nd} and \eqref{eqn:def-3rd},
 we obtain the integral of the third term in \eqref{eqn:ABI_eq98'}
\begin{multline}
\oint_{\Gamma_{m}}
\beta^{2}\left(\dfrac{X_{\infty}-X_{0}}{Y_{\infty}}\right)^{2}\left(
\dfrac{Y_{\infty}}{X-X_{\infty}}\right)^{2}\dfrac{dX}{Y}\\
=\dfrac{1}{2z}\left\{
-4\eta_{1}-4\omega_{1}\left(
\dfrac{x}{3}-2\beta^{2}z
\right)+\dfrac{1}{2}\left\{\pi(x_{+})+\pi(x_{-})\right\}
+2a\pi i
\right\}\,. \label{eqn:int-3-3}
\end{multline}
Thus, by using \eqref{eqn:int-3-2} and \eqref{eqn:int-3-3}, we conclude
\begin{equation}\label{eqn:ABI_eq108}
\mathcal{I}_{2}(\Gamma_{m})
=-\dfrac{1}{z}\left[
\eta_{1}+\omega_{1}\cdot\dfrac{x}{3}-\dfrac{1}{8}\dfrac{v}{\sqrt{z}}\left\{
\pi(x_{+})+\pi(x_{-})+2a\pi i
\right\}
\right]\,.
\end{equation}

We return to the calculation of $F_{1}$.
Using \eqref{eqn:ABI_eq96} and \eqref{eqn:ABI_eq108}, we have
\begin{equation}
v\mathcal{I}_{1}(\Gamma_{m})-z\mathcal{I}_{2}(\Gamma_{m})-\dfrac{\pi i}{4}\dfrac{v}{\sqrt{z}}
=\eta_{1}+\omega_{1}\cdot {x \over 3}+\dfrac{v}{8\sqrt{z}}\{\pi(x_{+})+\pi(x_{-})\}+\dfrac{\pi i}{4}(a-1)\cdot\dfrac{v}{\sqrt{z}}\,.
\end{equation}
Thanks to \eqref{eqn:ypm}, we find that the integrands of
\begin{equation}
\pi(x_{+})
=-2\int_{e_{3}}^{e_{2}}
\dfrac{y_{+}}{X-x_{+}}\dfrac{dX}{Y},
\quad
\pi(x_{-})
=-2\int_{e_{3}}^{e_{2}}
\dfrac{y_{-}}{X-x_{-}}\dfrac{dX}{Y}\,,
\end{equation}
are pure imaginary and real, respectively, so that we have
\begin{equation}
\pi(x_{+})\in i\mathbb{R},
\quad
\pi(x_{-})\in\mathbb{R}\,.
\end{equation}
Using this fact, we find
\begin{multline}\label{eqn:ABI_eq113}
v\mathcal{I}_{1}(\Gamma_{m})-z\mathcal{I}_{2}(\Gamma_{m})-\dfrac{\pi i}{4}\dfrac{v}{\sqrt{z}}
+\text{c.c}\\
=2\eta_{1}+\omega_{1}\cdot\dfrac{2x}{3}+\dfrac{1}{4}\{iv_{+}\pi(x_{+})+v_{-}\pi(x_{-})\}-\dfrac{\pi i}{2}(a-1)v_{+}\,.
\end{multline}
Therefore, substituting \eqref{eqn:ABI_eq72} and \eqref{eqn:ABI_eq113} into
\eqref{eqn:ABI_eq41}, we obtain
\begin{equation}\label{eqn:ABI_eq114}
F_{1}=-8\eta_{1}+8(x_{+}+x_{-})\omega_{1}
-\left\{
iv_{+}\pi(x_{+})+v_{-}\pi(x_{-})
\right\}+2\pi(a-1)v_{+}\,.
\end{equation}
Here, we have used $x=6(x_{+}+x_{-})$.
Thus, the final result for $F=F_{2}+F_{1}$ is
\begin{equation}\label{eqn:ABI_eq115}
F=-8\eta_{1}+\left(8\omega_{1}-\dfrac{3}{2h}\right)
(x_{+}+x_{-})
-\left\{
iv_{+}\pi(x_{+})+v_{-}\pi(x_{-})
\right\}+2\pi(a-1)v_{+}\,.
\end{equation}

\subsubsection{Derivation of the Atiyah-Hitchin metric}
The K\"ahler potential of the Atiyah-Hitchin manifold is obtained by the following generalized Legendre transformation:
\begin{equation}\label{eqn:ABI_eq116}
K(u,\bar{u},z,\bar{z})
=F(x,z,\bar{z},v,\bar{v})-(uv+\bar{u}\bar{v})\,,
\end{equation}
with the conditions \eqref{Fv} and \eqref{Ft}.
For the Atiyah-Hitchin manifold, they are reduced to
\begin{eqnarray}\label{eqn:dFdv=v_dFdx=0}
\dfrac{\partial F}{\partial v}=u\,,\quad
\dfrac{\partial F}{\partial x}=0\,.
\end{eqnarray}
Here the K\"ahler metric satisfies the hyperk\"ahler Monge-Amp\`ere equation \eqref{eqn:ma}.
Let us consider the first condition in \eqref{eqn:dFdv=v_dFdx=0}.
This yields
\begin{equation}\label{eqn:ABI_121}
u=-\dfrac{1}{2}\cdot\dfrac{1}{\sqrt{z}}\{\pi(x_{+})+\pi(x_{-})\}-\dfrac{\pi i}{\sqrt{z}}(a-1)\,.
\end{equation}
From this, we find
\begin{equation}\label{eqn:ABI_eq122}
uv+\bar{u}\bar{v}
=-\left\{iv_{+}\pi(x_{+})+v_{-}\pi(x_{-})\right\}+2(a-1)\pi v_{+}\,.
\end{equation}
On the other hand, 
since we find
\begin{equation}\label{eqn:dfdx=homega1}
\dfrac{\partial F}{\partial x}
=-\dfrac{1}{h}+4\omega_{1}\,,
\end{equation}
the second condition in \eqref{eqn:dFdv=v_dFdx=0} gives
\begin{equation}\label{eqn:ABI_eq123}
\dfrac{1}{h}=4\omega_{1}\,.
\end{equation}
Substituting
\eqref{eqn:ABI_eq115}, \eqref{eqn:ABI_eq122}, and \eqref{eqn:ABI_eq123} into
\eqref{eqn:ABI_eq116}, we find
\begin{equation}\label{eqn:ABI_eq125}
K=-8\eta_{1}+2(x_{+}+x_{-})\omega_{1}\,.
\end{equation}

From this K\"ahler potential, we shall derive the K\"ahler metric.
We introduce the holomorphic coordinate $U, Z$:
\begin{equation}\label{eqn:uz_UZ}
U=u\sqrt{z}\,, \quad
Z=2\sqrt{z}\,.
\end{equation}
The holomorphic coordinate change $(u, z)\mapsto(U, Z)$ preserves hyperk\"ahler Monge-Amp\`ere equation \eqref{eqn:ma}, namely,
\begin{equation}\label{eqn:ABI_eq127}
\det\left(
\begin{array}{cc}
K_{U\bar{U}} & K_{U\bar{Z}} \\
K_{Z\bar{U}} & K_{Z\bar{Z}}
\end{array}
\right)=1\,.
\end{equation}
In the following, we calculate $K_{Z\bar{Z}}$, $K_{Z\bar{U}}$, $K_{U\bar{Z}}$, $K_{U\bar{U}}$.
From \eqref{eqn:ABI_121}, we have
\begin{align}
U &= -\dfrac{1}{2}\left\{\pi(x_{+})+\pi(x_{-})\right\}-\pi i(a-1)\,,\\
\bar{U}  &= -\dfrac{1}{2}\left\{-\pi(x_{+})+\pi(x_{-})\right\}+\pi i(a-1)\,.
\end{align}
Their derivatives are
\begin{align}
dU &= -\dfrac{1}{2}\left\{d\pi(x_{+})+d\pi(x_{-})\right\}\,,\label{eqn:dU}\\
d\bar{U}  &= -\dfrac{1}{2}\left\{-d\pi(x_{+})+d\pi(x_{-})\right\}\,.\label{eqn:dUbar}
\end{align}
Here, it is shown that $d\pi(x_\pm)$ have the following form
\begin{equation}\label{eqn:ABI_eq130}
d\pi(x_{\pm})
=4A_\pm dx_{\pm}
+\dfrac{8(x_{\pm}^{2}-V\eta_{1})}{y_{\pm}}d\omega_{1}-8B_\pm d\eta_1\,,
\end{equation}
where
\begin{equation}
A_{\pm} = \dfrac{x_{\pm}\omega_{1}+\eta_{1}}{y_{\pm}}\,,\quad
B_{\pm} = \dfrac{x_{\pm}+V\omega_{1}}{y_{\pm}}\,, \quad
V=\dfrac{-3g_{3}\omega_{1}+2g_{2}\eta_{1}}{12\eta_{1}^{2}-g_{2}\omega_{1}^{2}}\,.
\end{equation}
Its detailed derivation is given in Appendix C.2 in \cite{Arai:2022xyc}.
From \eqref{eqn:ABI_eq123}, we find $d\omega_{1}=0$.
Therefore, $d\pi(x_\pm)$ reduces to
\begin{equation}
d\pi(x_{\pm})=4A_{\pm}dx_{\pm}-8B_{\pm}d\eta_{1}\,.
\end{equation}
Therefore, we obtain
\begin{equation}\label{eqn:ABI_eq134+135}
dU\mp d\bar{U}=-4A_{\pm}dx_{\pm}+8B_{\pm}d\eta_{1}\,.
\end{equation}
From $|Z|^{2}=x_{+}-x_{-}$, we have
\begin{equation}\label{eqn:ABI_eq136}
\bar{Z}dZ+Zd\bar{Z}=dx_{+}-dx_{-}\,.
\end{equation}
Utilizing \eqref{eqn:ABI_eq134+135} and \eqref{eqn:ABI_eq136}, we have the following relation:
\begin{eqnarray}
A_{-}(dU-d\bar{U})-A_{+}(dU+d\bar{U})
=-4A_{+}A_{-}(\bar{Z}dZ+Zd\bar{Z})+8(A_{-}B_{+}-A_{+}B_{-})d\eta_{1}\,.
\end{eqnarray}
We obtain $d\eta_{1}$ by solving this as
\begin{eqnarray}\label{eqn:deta_UZ}
 d\eta_{1}=\dfrac{A_{-}(dU-d\bar{U})-A_{+}(dU+d\bar{U})+4A_{+}A_{-}(\bar{Z}dZ+Zd\bar{Z})}{8(A_{-}B_{+}-A_{+}B_{-})}\,.\label{eqn:ABI_eq138}
\end{eqnarray}
Similarly, we have
\begin{equation}
(-B_{+}+B_{-})dU-(B_{+}+B_{-})d\bar{U}
=4(A_{-}B_{+}-A_{+}B_{-})dx_{\pm}-4A_{\mp}B_{\pm}(\bar{Z}dZ+Zd\bar{Z})\,,
\end{equation}
from which, we find $dx_\pm$ as
\begin{eqnarray}\label{eqn:dxpm_UZ}
dx_{\pm}
=\dfrac{(-B_{+}+B_{-})dU-(B_{+}+B_{-})d\bar{U}+4A_{\mp}B_{\pm}(\bar{Z}dZ+Zd\bar{Z})}{4(A_{-}B_{+}-A_{+}B_{-})}\,.\label{eqn:ABI_eq140+141}
\end{eqnarray}

We are ready to calculate the components of the K\"ahler metric.
Differentiating \eqref{eqn:ABI_eq125} with respect to $Z$, we find
\begin{align}
K_{Z} &= 2\bar{Z}\dfrac{-2A_{+}A_{-}+(A_{-}B_{+}+A_{+}B_{-})\omega_{1}}{A_{-}B_{+}-A_{+}B_{-}} \notag \\
&=-\dfrac{2}{Z}\left\{2\eta_{1}+(x_{+}+x_{-})\omega_{1}\right\}\,.
\end{align}
Further differentiating this with respect to $\bar{Z}$, we obtain $K_{Z\bar{Z}}$.
Similarly, we can calculate $K_{U\bar{Z}}$ and $K_{Z\bar{U}}$.
The component $K_{U\bar{U}}$ is obtained by solving \eqref{eqn:ma} with respect to $K_{U\bar{U}}$.
The components are explicitly given by
\begin{align}
K_{Z\bar{Z}}&=
-\dfrac{2(A_{+}A_{-})+2(A_{-}B_{+}+A_{+}B_{-})\omega_{1}}{A_{-}B_{+}-A_{+}B_{-}}\,,\label{eqn:ABI_eq143'}\\
K_{U\bar{Z}}&=-\dfrac{1}{2\bar{Z}}\dfrac{A_{-}-A_{+}+2(-B_{+}+B_{-})\omega_{1}}{A_{-}B_{+}-A_{+}B_{-}}\,,\label{eqn:ABI_eq144'}\\
K_{Z\bar{U}}&=\dfrac{1}{2Z}\dfrac{A_{-}+A_{+}+2(B_{+}+B_{-})\omega_{1}}{A_{-}B_{+}-A_{+}B_{-}}\,,\label{eqn:ABI_eq145'}\\
K_{U\bar{U}}&=\dfrac{1}{K_{Z\bar{Z}}}(1+K_{Z\bar{U}}K_{U\bar{Z}})\,.\label{eqn:ABI_eq146'}
\end{align}

As explained in Introduction, the Atiyah-Hitchin manifold has $SO(3)$-isometry.
In Section \ref{sec:AH_SLag}, we consider its Lie subgroup $SO(2)$ and construct a special Lagrangian submanifold 
 invariant under that symmetry.

\section{Special Lagrangian submanifold in Calabi-Yau manifold}\label{sec:SLS}

In this section, we review on special Lagrangian submanifolds 
 in a Calabi-Yau manifold and explain our method to construct special Lagrangian submanifolds
with a large symmetry, which was originally given by
Joyce \cite{joyce}.
We first recall the definition of special Lagrangian submanifolds in a general setting.
Next, we explain that any special Lagrangian submanifold
is a calibrated submanifold \cite{HL}.
From this, we obtain an equivalence condition that any Lagrangian submanifold
becomes a special Lagrangian submanifold.
Finally, we review the moment map approach to construct
special Lagrangian submanifolds in a Calabi-Yau manifold
in the case when the Calabi-Yau manifold
admits a Lie group which acts on it Hamiltonianly.
\subsection{Definition of special Lagrangian submanifold}
In this subsection, we recall the definition of special Lagrangian submanifolds in the context of Calabi-Yau geometry.

Let $(M,J, \omega,\Omega)$ be a complex $n$-dimensional Calabi-Yau manifold
and $g$ be the Riemannian metric on $M$ defined by
$g(X,Y)=\omega(X,JY)$ for any vector fields $X,Y$.
Here the K\"ahler form $\omega$ gives a closed non-degenerate $2$-form on $M$,
so that $(M,\omega)$ becomes a symplectic manifold.
Let $L$ be a Lagrangian submanifold of $M$ with respect to $\omega$.
Namely, $L$ is a real $n$-dimensional submanifold in $M$ satisfying
\begin{equation}\label{eqn:condition_Lag}
\omega|_{L}=0\,.
\end{equation}
The submanifold
$L$ has a Riemannian metric which is induced from the metric $g$ on $M$.
We denote by $d\mathrm{vol}_L$ is the volume form on $L$,
which gives an $n$-form on $L$.
The restriction of the complex volume form $\Omega$ to $L$,
which we write $\Omega|_L$, is also an $n$-form of $L$.
If $L$ is oriented, then the function on $L$ called the phase function $\theta: L \to S^1 = \mathbb{R}/2\pi\mathbb{Z}$ 
is defined by:
\begin{equation}
\Omega|_L = e^{-\sqrt{-1}\theta}d\mathrm{vol}_L\,.
\end{equation}
A de-Rham cohomology class of the closed $1$-form $d\theta$ is known as the Maslov class of $L$.
In this context, $L$ is called a special Lagrangian submanifold
if the phase function $\theta$ is constant,
namely, the de-Rham cohomology class of $d\theta$ vanishes.

\subsection{Special Lagrangian submanifold as calibrated submanifold}

In the context of Riemannian geometry,
Harvey and Lawson \cite{HL}
introduced the notion of calibrated submanifolds.
They also proved that
any special Lagrangian submanifold
is an example of calibrated submanifolds
in the case
when the ambient Riemannian manifold
has a Calabi-Yau structure.

\subsubsection{Calibrated submanifold}

In this subsection,
we first
recall the definition of 
calibrated submanifolds and give an equivalence condition that any Lagrangian submanifold
becomes a special Lagrangian submanifold
from the view point of the calibrated submanifolds.

Let $(M,g)$ be a Riemannian manifold.
A closed differential $n$-form $\eta$ on $M$
 is called a calibration if it satisfies the following inequality:
\begin{equation}
\eta(e_{1},\dotsc,e_{n})\leq 1\,, \label{eq:cal-ine}
\end{equation}
where $e_{1},\dotsc,e_{n}$ are (oriented) orthonormal vectors on $M$.
Then, for any
$n$-dimensional submanifold $L$ in $M$, the following relation holds:
\begin{equation}
\eta|_{L}\leq d\mathrm{vol}_{L}\,,
\end{equation}
where $\eta|_{L}$ is the restriction of $\eta$ to $L$. 
A calibrated submanifold $L$ in $M$
is defined as a submanifold
which attains the equality in \eqref{eq:cal-ine}
for all orthonormal bases $(e_{1},\dotsc,e_{n})$ of $T_{p}L$, $p\in L$,
equivalently,
\begin{equation}
\eta|_{L}=d\mathrm{vol}_{L}\,. \label{eq:cal-def}
\end{equation}

Let $(M,J,g,\Omega)$ be a Calabi-Yau manifold with complex dimension $n$.
By using a constant $\theta\in\mathbb{R}$, 
 we define a differential $n$-form on the Calabi-Yau manifold $M$ as $\eta\equiv\mathrm{Re}(e^{\sqrt{-1}\theta}\Omega)$.
It is known that this $\eta$ gives calibration on the Riemannian manifold $(M,g)$ and the corresponding 
 calibrated submanifold is a special Lagrangian submanifold with the phase $\theta$ \cite{HL}.
Furthermore, for a Lagrangian submanifold $L$,
  it can be shown that $L$
  is a special Lagrangian submanifold with phase $\theta$
  if and only if $L$ satisfies the following equation \cite{HL}
\begin{equation}\label{eqn:condition_SLag}
\mathrm{Im}(e^{\sqrt{-1}\theta}\Omega)|_{L}=0\,.
\end{equation}
In Section \ref{Taub-NUT_SLag}, we make use of this condition to construct special Lagrangian submanifolds
in the Taub-NUT manifold and in the Atiyah-Hitchin manifold.

\subsubsection{Homologically volume-minimality of calibrated submanifold}

The study of special Lagrangian submanifolds
 is attractive not only in Calabi-Yau geometry but also in Riemannian geometry.
In fact,
it is known that
any calibrated submanifolds
has a remarkable property called the homological volume-minimality.

Let $(M,g)$ be a Riemannian manifold
and $\eta\in\Omega^{n}(M)$ be a calibration on $M$.
We observe that any calibrated submanifold with respect to $\eta$
 has a volume minimality in the same homology class.
Namely, if $L$ is a calibrated submanifold in $M$ with respect to $\eta$,
 then $\mathrm{Vol}(L)\leq \mathrm{Vol}(L')$
 holds for any submanifold $L'$ which is homologous to $L$.
Here, $\mathrm{Vol}(L)$ and $\mathrm{Vol}(L')$ denote the volumes of $L$ and $L'$, respectively.
To verify this inequality, let $P$ be an $(n+1)$-dimensional submanifold with boundary $\partial P\equiv L\cup (-L')$.
According to Stokes' theorem, we find
\begin{equation}
0=\int_{P}d\eta
=\int_{\partial P}\eta
=\int_{L}\eta-\int_{L'}\eta\,.
\end{equation}
Thus, we have
\begin{equation}
\int_{L}\eta=\int_{L'}\eta\,,
\end{equation}
from which the following relation holds:
\begin{equation}
\mathrm{Vol}(L)
=\int_{L}d\mathrm{vol}_{L}
=\int_{L}\eta
= \int_{L'}\eta
\leq 
\int_{L'}d\mathrm{vol}_{L'}
=\mathrm{Vol}(L')\,.
\end{equation}
Here, the second equality follows from that $L$ is a calibrated submanifold, and the inequality
 follows from that $\eta$ is a calibration.
It has been shown that $L$ is a submanifold with a minimal volume
 in the same homology class.
In particular, a special Lagrangian submanifold has also the same property in the case when $M$ is a Calabi-Yau manifold.

\subsection{Review on construction of Lagrangian submanifolds by moment map approach}\label{sec:SLag_const}

Let us consider the case when a Calabi-Yau manifold
admits a Lie group which acts on it Hamiltonianly.
In such a case,
Joyce \cite{joyce}
proposed a method
to construct special Lagrangian submanifolds
in the Calabi-Yau manifold,
which is called the moment map approach.
Then, the action defines the moment map on the Calabi-Yau manifold.
By this method,
we get special Lagrangian submanifolds with a large symmetry. 

More precisely,
in order to construct special Lagrangian submanifolds in the complex $n$-dimensional Calabi-Yau manifold $M$, 
 it is sufficient to give real $n$-dimensional submanifolds satisfying \eqref{eqn:condition_Lag} and \eqref{eqn:condition_SLag}.
The condition \eqref{eqn:condition_Lag} gives Lagrangian submanifolds while the condition \eqref{eqn:condition_SLag} constrains 
 it to be special Lagrangian submanifolds. 
By using the moment map,
 we rewrite the condition \eqref{eqn:condition_Lag}
 as shown in Subsection \ref{sec:moment_Lag}.
The obtained Lagrangian submanifolds
 have symmetry of cohomogeneity one.
Furthermore, we can solve \eqref{eqn:condition_SLag} with the use of cohomogeneity-one symmetry.
This property simplifies \eqref{eqn:condition_SLag}: This condition is generally given by a nonlinear partial differential
 equation, but it is reduced to an ordinary differential equation due to the cohomogeneity-one symmetry.
Thus, by solving the ordinary differential equation,
 we obtain special Lagrangian submanifolds in the Calabi-Yau manifold.

In this subsection, we start with reviewing
 some basic notions related to the moment map approach to construct Lagrangian submanifolds.

\subsubsection{Cohomogeneity-one action}

Let $M$ be a smooth manifold. 
Let $H$ be a compact connected Lie group.
We denote by $e$ the identity element in $H$.
A group action of $H$ on $M$ is defined as a smooth map
$\varphi: K \times M \to M$ such that
\begin{equation}
\varphi(g, \varphi(h,p)) = \varphi(gh, p), \quad \varphi(e,p) = p,
\quad p\in M\,, g,h\in H\,.
\end{equation}
In this case, for each $h \in H$, the map
\begin{equation}
\varphi_{h}: M \to M; \quad p \mapsto \varphi(h, p)
\end{equation}
is a diffeomorphism on $M$.
We denote $\varphi(h,p)$ simply as $h \cdot p$. 
For any element $p \in M$, the set $H\cdot p:=\{h\cdot p\mid h\in H\}$ is called the $H$-orbit through $p$. 
The subgroup $H_{p}:=\{h \in H \mid h \cdot p = p\}$
 of $H$ is called the isotropy subgroup at $p$.
We also write the conjugacy class of $H_{p}$ as $[H_{p}]:=\{h H_{p} h^{-1} \mid h \in H\}$.
Here, $h H_{p} h^{-1}$ coincides with the isotropy subgroup $H_{h\cdot p}$ at the point $h\cdot p$.
For two points $p, q \in M$, their $H$-orbits satisfy either $H \cdot p = H \cdot q$ or $(H \cdot p) \cap (H \cdot q) = \emptyset$. 
Therefore, we can define an equivalence relation based on whether two points $p, q \in M$ belong to the same $H$-orbit. 
The quotient space of $M$ under this equivalence relation is called the orbit space, which is denoted as $H\backslash M=\{H\cdot p \mid  p\in M\}$.

Among the $H$-orbits,
the one with the maximum dimension is called the regular orbit, and those with dimensions lower than the regular orbit are called singular orbits. 
The $H$-action on $M$ is said to have 
 cohomogeneity one
if the codimension of the regular orbit in $M$ is equal to one. 
The orbit space of a cohomogeneity-one action becomes a 1-dimensional manifold (with boundaries).
Clearly,
in the case when $M$ is $m$-dimensional,
the dimensions of the regular orbits
of any cohomogeneity-one action on $M$
are equal to $m-1$.

\begin{ex}
The rotation of the sphere $S^{2}=\{x\in\mathbb{R}^{3}\mid \|x\|=1\}$ around the $z$-axis defines an $SO(2)$-action on $S^2$. 
Each $SO(2)$-orbit is determined by the intersection of $S^2$ and the plane $\{z=k\}$ parallel to the $xy$-plane. 
The orbits corresponding to $-1<k<1$ are the regular orbits, while the orbits corresponding to $k=\pm 1$ are the 
singular orbits. In particular, since the regular orbits are isomorphic to $S^1$, we find that this action is of cohomogeneity one.
The orbit space for the $SO(2)$-action on $S^{2}$ becomes $SO(2)\backslash S^2 \cong [-1,1]$.
\end{ex}

\subsubsection{Symplectic action and Hamiltonian action}

We first recall the definition of symplectic action.
Let $(M, \omega)$ be a symplectic manifold and $H$ be a compact connected Lie group.
An action $\varphi:H\times M\to M$ of $H$ on $(M,\omega)$
 is said to be \textit{symplectic}, if the following equation holds:
\begin{equation}\label{eqn:act_symplec'}
\varphi_{h}^{*}\omega = \omega\,,
\quad h\in H\,,
\end{equation}
that is, for any $p \in M$ and $X,Y\in T_{p}M$,
\begin{equation}\label{eqn:act_symplec}
\omega_{\varphi_{h}(p)}((\varphi_{h})_{*}X, (\varphi_{h})_{*}Y) = \omega_{p}(X, Y), \quad h\in H\,.
\end{equation}
Let $\mathfrak{h}$ denote the Lie algebra of $H$, and
$\exp: \mathfrak{h} \to H$ denote the exponential map of $H$. 
For each $X \in \mathfrak{h}$, the tangent vector field $X^{*}$ on $M$ is defined by:
\begin{equation}\label{eqn:X*_dfn}
X^{*}_{p} = \frac{d}{dt}\bigg|_{t=0}\varphi(\exp(tX), p), \quad p \in M\,,
\end{equation}
which is called the fundamental vector field associated with $X$. 
Then,
\eqref{eqn:act_symplec'}
yields
\begin{equation}\label{eqn:LXomega=0}
\mathcal{L}_{X^{*}}\omega = 0\,,
\end{equation}
for all $X\in\mathfrak{h}$, where $\mathcal{L}_{X^{*}}$ denotes
 the Lie derivative with respect to  $X^{*}$.
Furthermore, \eqref{eqn:LXomega=0} is equivalent that the $1$-form
 $\iota_{X^{*}}\omega$ is closed.
Here, $\iota_{X^{*}}:\Omega^{k}(M)\to \Omega^{k-1}(M)$ denotes the interior product of $X^{*}$:
\begin{equation}
\iota_{X^{*}}(\alpha)(X_{1}, \ldots, X_{k-1}) = \alpha(X^{*}, X_{1}, \ldots, X_{k-1})\,,
\quad
\alpha\in\Omega^{k}(M)\,,
X_{1},\dotsc,X_{k-1}\in\mathfrak{X}(M)\,.
\end{equation}

Let $\mathrm{Ad}:H\to\mathrm{GL}(\mathfrak{h})$
denote the adjoint representation of $H$.
Let $\mathfrak{h}^{*}$ denote
the dual vector space of $\mathfrak{h}$, that is,
the space of linear functionals on $\mathfrak{h}$.
We write $\INN{\cdot}{\cdot}:\mathfrak{h}^{*}\times\mathfrak{h}\to\mathbb{R}$ as the pairing
between $\mathfrak{h}^{*}$ and $\mathfrak{h}$.
The coadjoint representation $\mathrm{Ad}^*: H \to \mathrm{GL}(\mathfrak{h}^*)$ is defined by:
\begin{equation}
\langle \mathrm{Ad}^*(h)X, Y \rangle = \langle X, \mathrm{Ad}(h^{-1})Y \rangle, \quad h \in H, X \in \mathfrak{h}^*, Y \in \mathfrak{h}.
\end{equation}
For a symplectic action
$\varphi: H \times M \to M$ on $(M, \omega)$,
an $H$-equivariant map
$\mu:M\to \mathfrak{h}^{*}$
satisfying the following relation is called a moment map:
\begin{equation}\label{eqn:iotaxomega=dmux}
\iota_{X^{*}}\omega = d\mu_{X}\,,
\quad X\in\mathfrak{h}\,,
\end{equation}
where $\mu_{X}$
is a function on $M$
defined by
\begin{equation}\label{eqn:mu_mux}
\mu(p)(X) = \mu_{X}(p)\,,
\quad
p\in M\,,
X\in\mathfrak{h}\,.
\end{equation}
Here, the $H$-equivariance of the moment map $\mu$
means
\begin{equation}
\mu(h \cdot p) = \mathrm{Ad}^*(h)(\mu(p)), \quad p \in M, h \in H.
\end{equation}
A symplectic action $\varphi: H \times M \to M$ on $(M, \omega)$ is called a 
Hamiltonian action if there exists a moment map $\mu:M\to\mathfrak{h}^{*}$.
\begin{ex}\label{ex:R3_paratrans_Ham}
Translation and rotation within $\mathbb{R}^3$ are examples of Hamiltonian actions. 
Through this example, we can understand that the moment map is related to momentum and angular momentum, respectively.

We identify the cotangent bundle $T^{*}\mathbb{R}^{3}$ of $\mathbb{R}^{3}$ with $\mathbb{R}^{6}=\{(q,p)\mid q,p\in\mathbb{R}^{3}\}$. 
On $T^{*}\mathbb{R}^{3}=\mathbb{R}^{6}$, the standard symplectic form is defined as follows:
\begin{equation}
\omega=dq^{1}\wedge dp^{1}+dq^{2}\wedge dp^{2}+dq^{3}\wedge dp^{3}.
\end{equation}
The group $H=SO(3)=\{A\in \mathbb{R}^{3\times 3} \mid {}^{t}AA=E_{3}\}$ represents rotations around the origin in $\mathbb{R}^{3}$. 
We extend its action on $\mathbb{R}^{3}$ to that on $T^{*}\mathbb{R}^{3}=\mathbb{R}^{6}$, namely,
\begin{equation}
\varphi: SO(3)\times \mathbb{R}^{6} \to \mathbb{R}^{6};~ (A,(q,p)) \mapsto (Aq, Ap)\,.
\end{equation}
In what follows, we confirm that the action $\varphi$ is a Hamiltonian action.

Let $J_{i}$ ($i=1,2,3$) denote the element in the Lie algebra $\mathfrak{h}=\mathfrak{so}(3)$
defined by
\begin{equation}
J_{1}
=\left(
\begin{array}{ccc}
0 & 0 & 0 \\
0 & 0 & 1 \\
0 & -1 & 0 
\end{array}
\right)\,,\quad
J_{2}
=\left(
\begin{array}{ccc}
0 & 0 & 1 \\
0 & 0 & 0 \\
-1 & 0 & 0 
\end{array}
\right)\,,\quad
J_{3}
=\left(
\begin{array}{ccc}
0 & -1 & 0 \\
1 & 0 & 0 \\
0 & 0 & 0 
\end{array}
\right)\,,
\end{equation}
which consist of a basis of $\mathfrak{h}$.
Then we identify $\mathfrak{so}(3)$ with $\mathbb{R}^{3}$
as vector spaces via $J_{i}\mapsto e_{i}$,
where $\{e_{i}\}_{i=1,2,3}$
denotes the standard basis of $\mathbb{R}^{3}$.
For an element $a=a^{1}J_{1}+a^{2}J_{2}+a^{3}J_{3}=(a^{1},a^{2},a^{3})$
in the Lie algebra $\mathfrak{h}$,
the basic vector field $a^{*}$
can be expressed using the cross product $\times$ on $\mathbb{R}^{3}$ as follows:
\begin{align}
a^{*}_{(q,p)}
&= (a\times q, a\times p) \notag \\
&=(a^{2}q^{3}-a^{3}q^{2},-a^{1}q^{3}+a^{3}q^{1},a^{1}q^{2}-a^{2}q^{1},\notag\\
&\phantom{=}\quad \quad\quad\quad a^{2}p^{3}-a^{3}p^{2},-a^{1}p^{3}+a^{3}p^{1},a^{1}p^{2}-a^{2}p^{1})\,.
\end{align}
Hence the interior product $\iota_{a^{*}}\omega$ can be written as:
\begin{align}
\iota_{a^{*}}\omega \notag
&=\iota_{(a\times q,a\times p)}(dq^{1}\wedge dp^{1}+dq^{2}\wedge dp^{2}+dq^{3}\wedge dp^{3}) \notag \\
&=a^{1}(q^{2}dp^{3}+p^{3}dq^{2}-q^{3}dp^{2}-p^{2}dq^{3})\notag\\
&\phantom{=}\quad \quad+a^{2}(q^{3}dp^{1}+p^{1}dq^{3}-q^{1}dp^{3}-p^{3}dq^{1})\notag\\
&\phantom{=}\quad \quad+a^{3}(q^{1}dp^{2}+p^{2}dq^{1}-q^{2}dp^{1}-p^{1}dq^{2})\,.\label{eqn:moment1}
\end{align}
On the other hand,
we define
the mapping $ \mu: \mathbb{R}^{6}\to \mathfrak{h}^{*}=\mathfrak{so}(3)^{*}$ as follows:
\begin{align}
\mu(q,p)(a)
&=\INN{q\times p}{a} \notag \\
&=a^{1}(q^{2}p^{3}-q^{3}p^{2})+a^{2}(-q^{1}p^{3}+q^{3}p^{1})+a^{3}(q^{1}p^{2}-q^{2}p^{1})\,,\label{eqn:moment2}
\end{align}
for $(q,p)\in\mathbb{R}^{6}$,
$a=(a^{1},a^{2},a^{3})\in\mathbb{R}^{3}\cong\mathfrak{h}$,
where $\INN{\cdot}{\cdot}$ denotes the standard inner product on $\mathbb{R}^{3}$.
Then it follows from \eqref{eqn:moment1} and \eqref{eqn:moment2}
that $d\mu_{a}=\iota_{a^{*}}\omega$ holds.
We denote by $\{J^{i}\}_{i=1,2,3}$
the dual basis of $\mathfrak{h}^{*}=\mathfrak{so}(3)^{*}$
for the above basis $\{J_{i}\}_{i=1,2,3}$.
The correspondence $J^{i}\mapsto e_{i}$
yields an identification between $\mathfrak{h}^{*}=\mathfrak{so}(3)^{*}$ and $\mathbb{R}^{3}$.
Hence we have
\begin{align}
\mu(A\cdot (q,p))(a)
= \mu(Aq,Ap)(a) 
=\INN{A(q\times p)}{a}
=(\mathrm{Ad}^{*}(A)\mu(q,p))(a)\,,
\end{align}
for all $A\in SO(3)$.
This shows the $SO(3)$-equivariance of $\mu$.
Consequently, we have confirmed that the action $\varphi$ is Hamiltonian.
\end{ex}

\subsubsection{Moment map approach to construct Lagrangian submanifolds}\label{sec:moment_Lag}

The condition \eqref{eqn:condition_Lag} is nothing but the requirement for a submanifold to be isotropic.
We explain the isotropicity in terms of the moment map.
Our argument is based on \cite{HS}.

Let $(M, \omega)$ be a real $2n$-dimensional symplectic manifold
and $H$ be a compact connected Lie group
which acts Hamiltonianly on $(M,\omega)$.
We denote by
$\mu: M \to \mathfrak{h}^*$ the moment map associated with this Hamiltonian action.
Let $L$ be a connected, $H$-invariant submanifold in $M$.
The $H$-invariance of $L$
implies that
the fundamental vector fields
$X^{*}$
($X\in\mathfrak{h}$)
are tangent to the submanifold $L$.
In the case when $L$ is isotropic,
for any $X\in\mathfrak{h}$,
the smooth function $\mu_{X}\in C^{\infty}(M)$ defined in
\eqref{eqn:mu_mux} is constant on $L$ because
\begin{equation}
d\mu_{X}(Y)=\iota_{X^{*}}\omega(Y)=\omega(X^{*},Y)=0,
\quad Y\in\mathfrak{X}(L).
\end{equation}
Here, we have used the isotropicity of $L$
in the last equality.
Hence, by \eqref{eqn:mu_mux},
the moment map $\mu: M \to \mathfrak{h}^*$ is also constant on $L$,
that is,
there exists $c\in\mathfrak{h}^{*}$
such that $L$ is contained in the level set
$\mu^{-1}(c):=\{p \in M \mid \mu(p)=c\}$,
\begin{equation}\label{eqn:Lsubsetmuinverse}
L\subset \mu^{-1}(c)\,.
\end{equation}
Then,
for an element $p$ in $L$,
 we write $c=\mu(p)$.
For any $h\in H$,
by the $H$-equivariance of $\mu$,
we have
\begin{equation}
\mathrm{Ad}^{*}(h)c
=\mathrm{Ad}^{*}(h)\mu(p)
=\mu(h\cdot p).
\end{equation}
Since $L$ is $H$-invariant,
we have $\mu(h\cdot p)=c$,
so that we obtain $\mathrm{Ad}^{*}(h)c=c$.
This means that
the element $c$
is contained in
the subset
$Z(\mathfrak{h}^*) := \{\xi \in \mathfrak{h}^* \mid \mathrm{Ad}^*(h) \xi = \xi, \, h \in H\}$ of $\mathfrak{h}^{*}$.

Conversely,
for any connected, $H$-invariant submanifold
$L$ in $M$
satisfying \eqref{eqn:Lsubsetmuinverse}
for some $c\in Z(\mathfrak{h}^{*})$,
it is shown that,
if the $H$-action on $L$ induced from that on $M$
is cohomogeneity one, then
$L$ becomes an isotropic submanifold in $(M,\omega)$.
Indeed,
\eqref{eqn:Lsubsetmuinverse} yields
\begin{equation}\label{eqn:omegaXY_zero}
\omega(X^{*},Y)=d\mu_X(Y)=0\,,\quad
X\in\mathfrak{h},Y\in\mathfrak{X}(L).
\end{equation}
Let $p\in L$ be an element in $L$ such that
the $H$-orbit through this point is regular.
The space $\mathrm{span}_{\mathbb{R}}\{X^{*}_{p}\mid
X\in\mathfrak{h}\}$ gives a codimension one subspace of $T_{p}L$.
Let $v\in T_{p}L$ be a non-zero vector which is orthogonal to $\mathrm{span}_{\mathbb{R}}\{X^{*}_{p}\mid
X\in\mathfrak{h}\}$.
Since any tangent vector $Y_{p}\in T_{p}L$
is decomposed into $Y_{p}=X^{*}+a\,v$
for some $X\in \mathfrak{h}$
and $a\in\mathbb{R}$,
we have 
\begin{equation}
\omega_{p}(v,Y_{p})=\omega_{p}(v,X^{*})+a\,\omega_{p}(v,v)=0.
\end{equation}
This implies that $\omega_{p}$ is isotropic at $p\in L$.
By the arbitrariness of $p$,
$\omega$ vanishes on a open subset of $L$.
By the connectedness of $L$,
$\omega$ vanishes on the whole $L$,
and therefore $L$ is isotropic.

From the above argument,
in order to obtain Lagrangian submanifolds,
it is sufficient to construct
connected, $H$-invariant submanifolds
$L$ in $M$ satisfying 
\eqref{eqn:Lsubsetmuinverse}
and $2\dim(L)=\dim(M)$.

\subsection{Construction of special Lagrangian submanifolds by using moment map}\label{sec:moment_SLag_const}
Let $(M, J, \omega, \Omega)$ be a Calabi-Yau manifold.
Let $H$ be a compact connected Lie group which acts Hamiltonianly on $M$.
Then, we write $\mu:M\to \mathfrak{h}^{*}$
as the moment map associated with this Hamiltonian action.
As discussed in Sections \ref{Taub-NUT_SLag} and
\ref{sec:AH_SLag},
we may assume that $M$ has dimension $4$,
and that $H$ is a $1$-dimensional Lie subgroup of the isometry group of $M$.

First, based on the argument in Subsection \ref{sec:moment_Lag}, we explain the method for constructing Lagrangian submanifolds in $(M,\omega)$. 
Let $c \in Z(\mathfrak{h}^{*})$.
We consider a curve
\begin{equation}
\ell: I \to M; \quad t \mapsto \ell(t),
\end{equation}
in $M$ such that its image is contained in $\mu^{-1}(c)$, and at each point, the curve intersects the $H$-orbit passing through that point transversally. 
The later condition guarantees that
\begin{equation}
L = H \cdot \ell := \{h \cdot \ell(t) \mid h \in H, t \in I\},
\end{equation}
gives a $2$-dimensional (i.e., half dimensional) submanifold in $M$. 
By the $H$-equivariance of the moment map $\mu$,
the former condition implies that $L$ is contained in $\mu^{-1}(c)$.
Since the action of $H$ on $L$ has cohomogeneity one,
 $L$ becomes a Lagrangian submanifold as shown in Subsection \ref{sec:moment_Lag}.

Next, we find special Lagrangian submanifolds among such Lagrangian submanifolds $L$, i.e., those 
 satisfying the condition \eqref{eqn:condition_SLag}. 
If the condition \eqref{eqn:condition_SLag} is satisfied at a point $p$ of $L$, then it is satisfied at every point of 
 the orbit $H \cdot p$. 
Therefore, it is sufficient to demonstrate that the condition is satisfied at the point $\ell(t)$
on the curve $\ell$ for each $t \in I$.
Since $H$ is 1-dimensional, it can be expressed as $H=\{\exp(tX) \mid t \in \mathbb{R}\}$ for some non-zero 
 $X \in \mathfrak{h}$. 
In this case, the tangent space $T_{\ell(t)}L$ of $L$ at the point $\ell(t)$ is spanned by
$X^{*}_{\ell(t)}$ and the velocity vector $\dot{\ell}(t)$ of the curve $\ell$
 since $L$ has cohomogeneity one.
Therefore, the condition \eqref{eqn:condition_SLag} can be expressed as:
\begin{equation}\label{eqn:condition_SLag_curve}
\mathrm{Im}(e^{\sqrt{-1}\theta}\Omega(\dot{\ell}(t),X^{*}_{\ell(t)}))=0\,.
\end{equation}

From the above argument,
we have constructed a special Lagrangian submanifold $L=H \cdot \ell$ for the phase $\theta$ by giving a solution curve $\ell$
for the ordinary differential equation \eqref{eqn:condition_SLag_curve}.

\section{Construction of special Lagrangian manifolds in the Taub-NUT manifold}\label{Taub-NUT_SLag}
In this section, we construct special Lagrangian submanifolds in the Taub-NUT manifold according to Subsections \ref{sec:SLag_const} and \ref{sec:moment_SLag_const}.

\subsection{The $U(1)$ tri-holomorphic isometry}

\subsubsection{Deriving the conditions for being special Lagrangian}
The following vector field $X$
generates a 1-parameter transformation group $H\cong U(1)$ that acts 
 Hamiltonianly on the Taub-NUT manifold $M$:
\begin{equation}
X=i\left(
\dfrac{\partial}{\partial u}-\dfrac{\partial}{\partial \bar{u}}
\right)\,.
\end{equation}
In fact, the following function
$\mu$ gives the corresponding moment map (up to constant):
 \begin{equation}\label{eqn:moment_U1}
\mu=\dfrac{1}{2}x\,.
\end{equation}
In order to verify this, since $U(1)$ is abelian,
 it is sufficient to show that $\iota_{X}\omega=d\mu$. 
To do this, we express $\iota_{X}\omega$ and $d\mu$ as linear combinations of $du,d\bar{u},dz$ and $d\bar{z}$ and 
 verify that each component is the same.
Direct calculation yields
\begin{equation}
\iota_{X}(du\wedge d\bar{u})=i(du+d\bar{u})\,, 
\iota_{X}(du \wedge d\bar{z}) = id\bar{z}\,, 
\iota_{X}(dz \wedge d\bar{u}) = idz\,, 
\iota_{X}(dz \wedge d\bar{z}) = 0\,.
\end{equation}
Hence,
by \eqref{eqn:kealer_K},
we have
\begin{align}
\iota_{X}\omega
&=\dfrac{i}{2}\left\{
K_{u\bar{u}}\iota_{X}(du\wedge d\bar{u})
+K_{u\bar{z}}\iota_{X}(du\wedge d\bar{z})
+K_{z\bar{u}}\iota_{X}(dz\wedge d\bar{u})
+K_{z\bar{z}}\iota_{X}(dz\wedge d\bar{z})
\right\} \notag\\
&=\dfrac{1}{2}F_{xx}^{-1}((du+d\bar{u})-F_{xz}dz-F_{x\bar{z}}d\bar{z})\,.\label{eqn:iotaXomega_comp}
\end{align}
Here, we have used 
\eqref{eqn:Kzzb_Fxx}--\eqref{eqn:Kzub_FuubFxx}
in the last equality.
On the other hand,
\eqref{eqn:Fxuub} yields
\begin{align}
dx &= \dfrac{\partial x}{\partial u}du + \dfrac{\partial x}{\partial \bar{u}}d\bar{u} + \dfrac{\partial x}{\partial z}dz + \dfrac{\partial x}{\partial \bar{z}}d\bar{z} 
 \notag \\
&= F_{xx}^{-1}\left((du + d\bar{u}) - F_{xz}dz - F_{x\bar{z}}d\bar{z}\right)\,.
\label{eqn:dx_F}
\end{align}
Thus, we obtain the moment map as \eqref{eqn:moment_U1}.

Following Subsection \ref{sec:moment_SLag_const}, we first
consider a curve in the Taub-NUT manifold $M$, which we write
\begin{equation}
\ell:I \to M;~t\mapsto \ell(t)
=(u(t),\bar{u}(t),z(t),\bar{z}(t))\,,
\end{equation}
where $I$ is an open interval in $\mathbb{R}$.
We set $L=H\cdot \ell$ and define
tangent vectors
$v_{1},v_{2}\in T_{\ell(t)}L$ as follows:
\begin{align}
v_{1} &= X_{\ell(t)}=i\left(\dfrac{\partial}{\partial u}-\dfrac{\partial}{\partial \bar{u}}\right)\,,\\
v_{2} &= \dot{\ell}(t)
=\dot{u}\dfrac{\partial}{\partial u}+\dot{\bar{u}}\dfrac{\partial}{\partial \bar{u}}+\dot{z}\dfrac{\partial}{\partial z}+\dot{\bar{z}}\dfrac{\partial}{\partial \bar{z}}\,.
\end{align}
In order to
guarantee that 
$\ell$ intersects transversally the $H$-orbit $H\cdot \ell(t)$
at $\ell(t)$ for each $t\in I$,
$\{v_{1},v_{2}\}$ must be linearly independent,
equivalently,
\begin{equation}
\mathrm{Re}(\dot{u}(t))\neq 0\,.
\end{equation}
Then $\{v_{1},v_{2}\}$
provides a basis for $T_{\ell(t)}L$.
From \eqref{eqn:moment_U1},
if $x$ takes a constant value $c\in\mathbb{R}\cong Z(\mathfrak{u}(1)^{*})$ on $\ell$,
then $L=H \cdot \ell$ is contained in $\mu^{-1}(c)$.
This means that $L$ becomes a Lagrangian submanifold in $M$.

Next, we rewrite the condition \eqref{eqn:condition_SLag_curve}
for $L$.
For simplicity, we may assume that the phase $\theta=0$.
Then, we have:
\begin{align}
\Omega(v_{1},v_{2})
&=du\wedge dz(v_{1},v_{2}) \notag \\
&=i\dot{z}(t)\,,
\end{align}
so that $ \mathrm{Im}(\Omega(v_{1},v_{2}))=0$ becomes:
\begin{equation}
\mathrm{Re}(\dot{z}(t))=0\,,
\end{equation}
equivalently,
\begin{equation}
\mathrm{Re}(z(t))=\text{constant}\,.
\end{equation}
From the above argument
we have obtained the conditions
for $\ell$
such that $L=H\cdot \ell$ is a special Lagrangian submanifold in $M$.
The conditions we obtained are the same ones derived in \cite{Noda} 
\footnote{Our definition of $\Omega$ is different
 from one in \cite{Noda} by $\sqrt{-1}$. 
 Considering this, our conditions and the conditions in \cite{Noda} are the same.}.
Note that we have derived them based on the generalized Legendre transform 
 approach together with the moment map technique.

\subsubsection{Analysis of the condition for being special Lagrangian}\label{sec:acSLS_TaubNUT}
In order for $L=H \cdot \ell$ to be a special Lagrange submanifold in the Taub-NUT manifold $M$,
the curve $\ell=(u, \bar{u}, z, \bar{z})$ must satisfy the following conditions:
\begin{equation}\label{eqn:reunotconst}
\mathrm{Re}(u) \neq \text{constant}\,,
\end{equation}
and
\begin{equation}\label{eqn:xconst_rezconst}
\quad
x = \text{constant}, \quad \mathrm{Re}(z) = \text{constant}.
\end{equation}
We would like to give curves $\ell$
satisfying
\eqref{eqn:reunotconst} and \eqref{eqn:xconst_rezconst}.

In what follows, we rewrite \eqref{eqn:xconst_rezconst}
in terms of special coordinates $(r,\theta,\phi,\psi)$,
and then obtain some solution curves of \eqref{eqn:xconst_rezconst} in the $(\phi, r)$-plane
and the $(\phi, \theta)$-plane, respectively.

We define:
\begin{equation}\label{eqn:zu_rtpp}
\begin{cases}
r \sin\theta \cos\phi = z + \bar{z}, \\
r \sin\theta \sin\phi = -i(z - \bar{z}), \\
r \cos\theta = x, \\
-2m\psi = \mathrm{Im}(u)\,.
\end{cases}
\quad (0 \le \theta \le \pi,~~0\le\phi\le 2\pi,~~0\le\psi \le 4\pi)
\end{equation}
Combined with \eqref{eqn:Fxuub}, the holomorphic coordinates $z$ and $u$ are expressed as follows:
\begin{equation}\label{eqn:TaubNUT_zu_rtpp}
z = \dfrac{1}{2}r\sin\theta e^{i\phi}, \quad
u = -2mi \psi -2\dfrac{r\cos\theta}{h} - 2m \log\dfrac{1+\cos\theta}{\sin\theta}\,.
\end{equation}
Hence \eqref{eqn:reunotconst} and \eqref{eqn:xconst_rezconst} are rewritten as:
\begin{equation}
\dfrac{r\cos\theta}{h} + m \log\dfrac{1+\cos\theta}{\sin\theta}
\neq \text{constant}\,,
\end{equation}
and
\begin{equation}\label{eqn:xc1_rezc2}
r \cos\theta = c_1\,,\quad
\dfrac{1}{2}r \sin\theta \cos\phi = c_2\,,
\end{equation}
where $c_1$ and $c_2$ are arbitrary constants.
Note that no conditions are imposed on $\psi$ for $L=H \cdot \ell$ to be a special Lagrangian
submanifold in the Taub-NUT manifold.

\paragraph{Case 1: The solution curves of \eqref{eqn:xc1_rezc2} in the $(\phi,r)$-plane}

\begin{figure}[H]
\centering
\begin{tabular}{cc}
\includegraphics[keepaspectratio, scale=0.5]{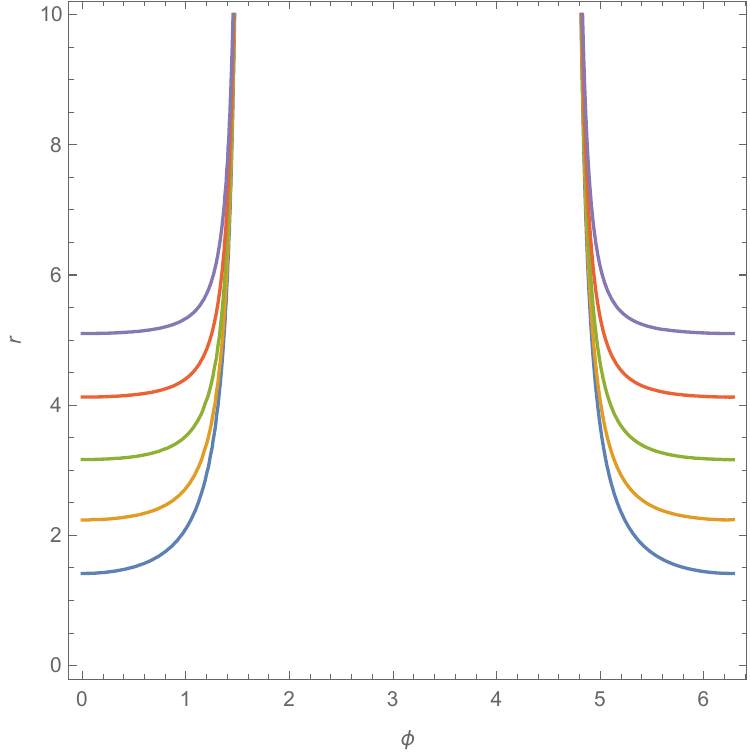}&
\includegraphics[keepaspectratio, scale=0.5]{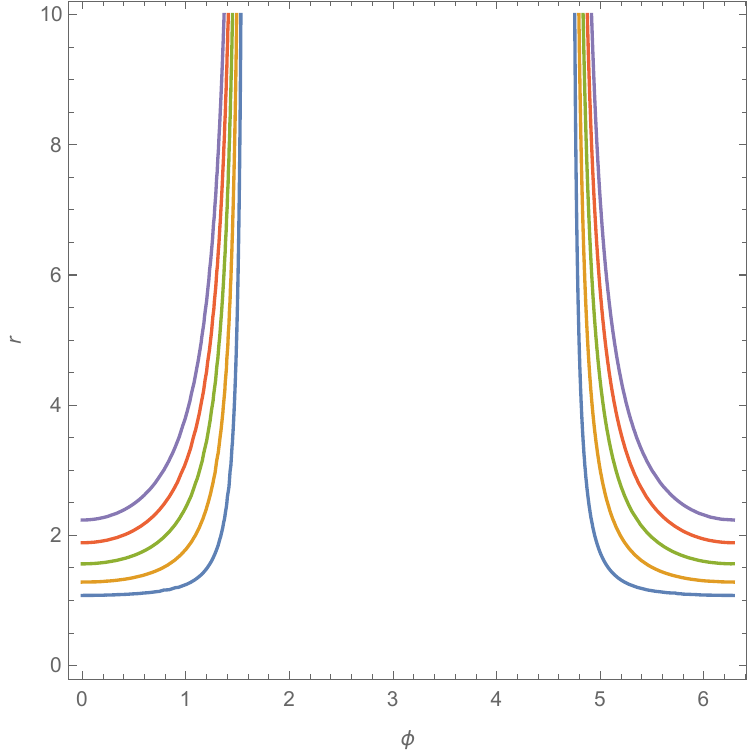}
\end{tabular}
\caption{Solution curves for 
$\cos\phi=\dfrac{2c_{2}}{\sqrt{r^{2}-c_{1}^{2}}}$ in the $(\phi,r)$-plane.}\label{fig:Taub_rphi}
\end{figure}

In order to eliminate $\theta$ from \eqref{eqn:xc1_rezc2},
we rewrite the first equation in \eqref{eqn:xc1_rezc2} as follows:
\begin{equation}\label{eqn:sin2theta=1-c12r2}
r^2 \sin^2 \theta = r^2 - c_1^2.
\end{equation}
Then, from the second equation in \eqref{eqn:xc1_rezc2}
we get
\begin{equation}
\dfrac{1}{4}(r^{2}-c_{1}^{2})\cos^{2}\phi=c_{2}^{2}\,,
\end{equation}
so that we have
\begin{equation}\label{eqn:SLS230216_92blw}
\cos \phi = \frac{2c_2}{\sqrt{r^2 - c_1^2}}.
\end{equation}
By giving constant values
$c_1$ and $c_2$, we obtain the solution curves of
\eqref{eqn:xc1_rezc2} in the $(\phi, r)$-plane.
In particular, such solution curves satisfy
\begin{equation}
\phi\to \dfrac{\pi}{2}\,,\,\dfrac{3}{2}\pi
\quad
\text{(as $r\to \infty$)}\,.
\end{equation}
In Figure \ref{fig:Taub_rphi}, we show the solution curves obtained from the specific values of $c_1$ and $c_2$. 
The left part of this figure shows the solution curves for $c_1 = 1, \ldots, 5$ and $c_2 = 1/2$,
and
the right part shows the solution curves for $c_1 = 1$ and $c_2 = 1/5, 2/5, \ldots, 1$. 
Then,
for each solution curve $\ell$,
the Lagrangian submanifold
$L = H \cdot \ell$ is a special Lagrangian submanifold in the Taub-NUT manifold $M$.

\paragraph{Case 2: The solution curves in the $(\phi,\theta)$-plane}

\begin{figure}[H]
\centering
\includegraphics[keepaspectratio, scale=0.5]{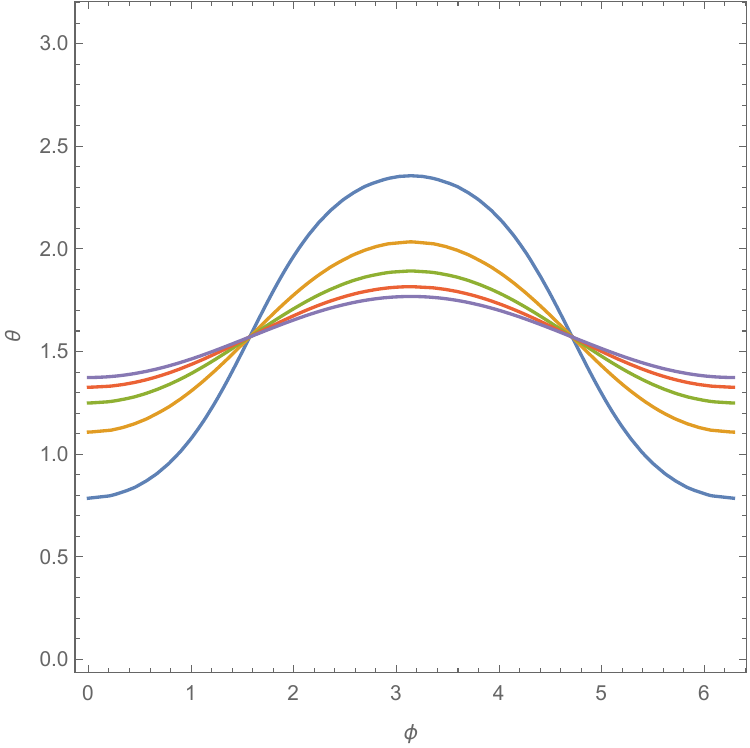}
\caption{Solution curves for 
$\cos\phi=\dfrac{c}{\tan\theta}$ in the $(\phi,\theta)$-plane.}\label{fig:Taub_thetaphi}
\end{figure}

Similar calculations as in Case 1 allow us to eliminate $r$ from \eqref{eqn:xc1_rezc2}. 
Namely, we obtain
\begin{align}
\cos\phi&=\dfrac{2c_{2}}{c_{1}}\cdot\dfrac{1}{\tan\theta}=\dfrac{c}{\tan\theta}\,,
\label{eqn:SLS230216_91}
\end{align}
where $c=2c_{2}/c_{1}$.
Thus, for each $c$, we have the solution curves of \eqref{eqn:xc1_rezc2}
in the $(\phi, \theta)$-plane. 
In Figure \ref{fig:Taub_thetaphi}, each curve corresponds to the solution curves of \eqref{eqn:SLS230216_91} for $c=1,2,\ldots,5$,
which represent special Lagrangian submanifolds in the Taub-NUT manifold $M$.

\subsection{The non-tri-holomorphic $SO(3)$ isometry}

\subsubsection{Deriving the conditions for being special Lagrangian}
We consider the vector field
\begin{equation}
X=-2i\left(
z\dfrac{\partial}{\partial z}-\bar{z}\dfrac{\partial}{\partial \bar{z}}
\right)\,,
\end{equation}
which generates a one-parameter transformation group $H\cong SO(2)$
 that acts Hamiltonianly on the Taub-NUT manifold. 
In fact,
the following function $\mu$
gives the corresponding moment map
(up to constant):
\begin{equation}\label{eqn:moment_SO2}
\mu =2mr+\dfrac{1}{2h}\cdot 4|z|^{2}\,.
\end{equation}
A similar argument in \eqref{eqn:moment_U1} yields \eqref{eqn:moment_SO2},
that is, $\iota_X\omega=d\mu$ holds.
Indeed,
by direct calculations, we have
\begin{equation}\label{eqn:moment_so2_iotaomega}
\iota_{X}\omega
=K_{u\bar{z}}\bar{z}du+K_{z\bar{u}}zd\bar{u}+K_{z\bar{z}}(\bar{z}dz+zd\bar{z})\,.
\end{equation}
On the other hand, 
by differentiating the both side of \eqref{eqn:moment_SO2},
we have
\begin{equation}
d\mu
=2m dr+\dfrac{1}{2h}(4\bar{z}dz+4zd\bar{z})\,.
\end{equation}
By differentiating the both sides of 
the second equation in
\eqref{eqn:zetapm_r},
we get
\begin{equation}\label{eqn:dr_taub_so3moment}
dr=\dfrac{x}{r}dx +2\dfrac{\bar{z}}{r}dz +2\dfrac{z}{r}d\bar{z}\,.
\end{equation}
From \eqref{Kzzb-TN}--\eqref{Kuub-TN},
we have
\begin{equation}
dx 
= -\dfrac{1}{2}V^{-1}du -\dfrac{1}{2}V^{-1}d\bar{u}+\dfrac{mx}{r|z|^{2}}V^{-1}\bar{z}dz
+\dfrac{mx}{r|z|^{2}}V^{-1}zd\bar{z}\,.
\end{equation}
Substituting this into \eqref{eqn:dr_taub_so3moment},
we have
\begin{equation}
dr
=-\dfrac{1}{2}\dfrac{x}{r}V^{-1}du-\dfrac{1}{2}\dfrac{x}{r}V^{-1}d\bar{u}\\
+\left(\dfrac{mx^{2}}{r^{2}|z|^{2}}V^{-1}+\dfrac{2}{r}\right)\bar{z}dz
+\left(\dfrac{mx^{2}}{r^{2}|z|^{2}}V^{-1}+\dfrac{2}{r}\right)zd\bar{z}\,.
\end{equation}
Thus, 
we obtain
\begin{align}
d\mu
&= -\dfrac{mx}{r}V^{-1}du-\dfrac{mx}{r}V^{-1}d\bar{u}
+\left(2\dfrac{m^{2}x^{2}}{r^{2}|z|^{2}}V^{-1}+\dfrac{4m}{r}+\dfrac{2}{h}\right)(\bar{z}dz+zd\bar{z})
\notag \\
&=K_{u\bar{z}}\bar{z}du+K_{z\bar{u}}zd\bar{u}+K_{z\bar{z}}(\bar{z}dz+zd\bar{z})\,,
\end{align}
which is equal to $\iota_X\omega$ as shown in \eqref{eqn:moment_so2_iotaomega}.

Following Subsection \ref{sec:moment_SLag_const},
we give special Lagrangian submanifolds with cohomogeneity-one symmetry $H$.
We first consider a curve $\ell$ in the Taub-NUT manifold $M$,
\begin{equation}
\ell:I \to M;~t\mapsto \ell(t)=(u(t),\bar{u}(t),z(t),\bar{z}(t))\,.
\end{equation}
We set $L=H\cdot \ell$ and define $v_{1},v_{2}\in T_{\ell(t)}L$ as follows:
\begin{align}
v_{1} &= X_{\ell(t)}=-2i\left(z\dfrac{\partial}{\partial z}-\bar{z}\dfrac{\partial}{\partial \bar{z}}\right)\,,\\
v_{2} &= \dot{\ell}(t)
=\dot{u}\dfrac{\partial}{\partial u}+\dot{\bar{u}}\dfrac{\partial}{\partial \bar{u}}+\dot{z}\dfrac{\partial}{\partial z}+\dot{\bar{z}}\dfrac{\partial}{\partial \bar{z}}\,.
\end{align}
In order to guarantee that $\ell$ intersects transversally the $H$-orbit $H\cdot \ell(t)$
at $\ell(t)$ for each $t$, $\{v_{1},v_{2}\}$ must satisfy
\begin{equation}
|z(t)| \neq \text{constant}\,.
\end{equation}
From  \eqref{eqn:moment_SO2},
for any constant value $c\in\mathbb{R}\cong Z(\mathfrak{so}(2)^{*})$,
if
\begin{equation}\label{eqn:HSO(2)TaubNUT_Lag}
2mr+\dfrac{1}{2h}\cdot 4|z|^{2} = c\,,
\end{equation}
then $L$ becomes a Lagrangian submanifold in $M$.

Next, we rewrite the condition \eqref{eqn:condition_SLag_curve}
for $L$ to be a special Lagrangian submanifold in $M$.
We calculate $\Omega(v_1, v_2)$:
\begin{align}
\Omega(v_{1},v_{2})
&=du\wedge dz(v_{1},v_{2}) \notag \\
&=2iz(t)\dot{u}(t)\,.
\end{align}
Hence \eqref{eqn:condition_SLag_curve} with $\theta=0$
is equivalent to
\begin{equation}\label{eqn:TaubNUT_SLS_SO3symm}
\mathrm{Re}(z(t)\dot{u}(t))=0\,.
\end{equation}
From the above argument, we have obtained the
conditions for $\ell$ such that
$L=H\cdot \ell$ is a special Lagrangian submanifold
in the Taub-NUT manifold $M$.

\subsubsection{Analysis of the condition for being special Lagrangian}

In a similar manner in Subsection \ref{sec:acSLS_TaubNUT},
we give some solution curves $\ell$ for 
\begin{equation}\label{eqn:TaubNUT_SLS_SO3}
2mr+\dfrac{1}{2h}\cdot 4|z|^{2}=\text{constant}\,,
\quad
\mathrm{Re}(z\dot{u})=0\,.
\end{equation}
The first equation in \eqref{eqn:TaubNUT_SLS_SO3} can be expressed by the spherical coordinates
\eqref{eqn:TaubNUT_zu_rtpp} as follows:
\begin{equation}\label{eqn:TaubNUT_SLS_SO31st'}
2mr+\dfrac{1}{2h}r^{2}\sin^{2}\theta=c_{1}\,,
\end{equation}
where $c_{1}$ is a constant.
On the other hand,
it is difficult to solve
the second equation in \eqref{eqn:TaubNUT_SLS_SO3} in general.
In what follows, we restrict our considerations to
the case where $\dot{u}(t)$ is a constant:
\begin{equation}\label{eqn:TaubNUT_SLS_SO32nd'}
\dot{u}(t)=c_{2}.\qquad
\text{($c_{2}$: constant)}
\end{equation}

\begin{figure}[H]
\centering
\includegraphics[keepaspectratio, scale=0.5]{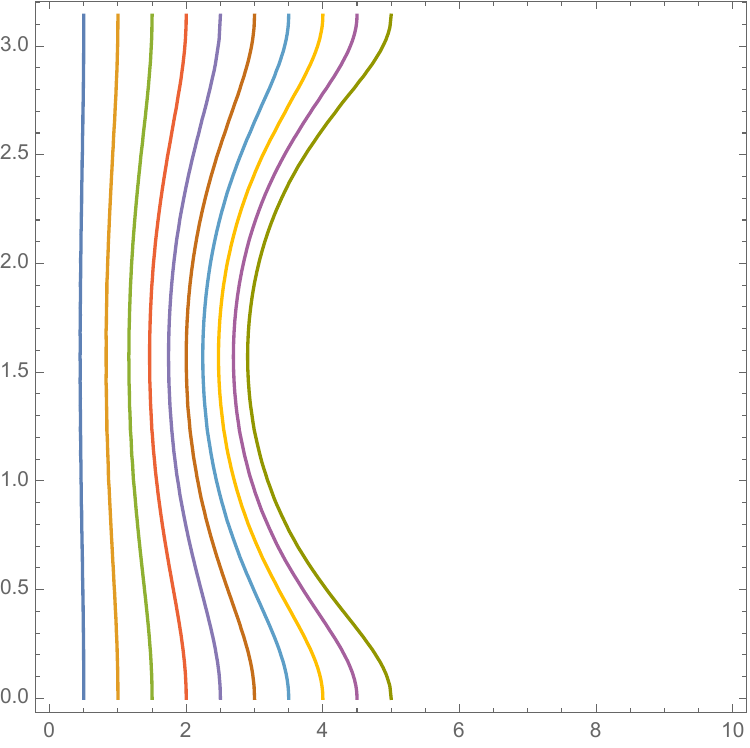}
\caption{Solution curves for 
$2mr+\dfrac{1}{2h}r^{2}\sin^{2}\theta=c_{1}$
in $(r,\theta)$-plane for $c_{1}=1,\dotsc,10$.}\label{fig:Taub_rtheta2}
\end{figure}

\paragraph{Case 1: $c_{2}$ is a real number}
In this case, $u$ takes real valued.
Hence, we have
\begin{equation}\label{eqn:c2real_psi=0}
\psi=0\,,
\end{equation}
by the fourth equation in \eqref{eqn:zu_rtpp}.
Also, the second equation in \eqref{eqn:TaubNUT_SLS_SO3}
yields
\begin{equation}
\mathrm{Re}(z)=0\,.
\end{equation}
By using \eqref{eqn:TaubNUT_zu_rtpp}
this equation is rewritten as follows:
\begin{equation}
r\sin\theta\cos\phi=0\,.
\end{equation}
Therefore, we get
\begin{equation}
\theta=0,\pi
\quad
\text{or}
\quad
\phi=\dfrac{\pi}{2},\dfrac{3}{2}\pi\,.
\end{equation}
Substituting this into \eqref{eqn:TaubNUT_SLS_SO31st'}
we have
\begin{equation}
\begin{cases}
2mr=c_{1} & (\theta=0\,,\,\pi)\,,\\
2mr+\dfrac{1}{2h}r^{2}\sin^{2}\theta=c_{1} & (\phi=\dfrac{\pi}{2}\,,\,\dfrac{3}{2}\pi)\,.
\end{cases}
\end{equation}
From the above arguments,
we obtain special Lagrangian submanifolds
$L=H\cdot \ell$ for the solution curves
$\ell(t)=(r(t),\theta(t),\phi(t),\psi(t))$,
\begin{equation}
\ell(t)=(c_{1}/2m,0,\phi(t),0)\,,\quad
(c_{1}/2m,\pi,\phi(t),0)\,,
\end{equation}
or
\begin{equation}
\ell(t)=(r(t),\theta(t),\dfrac{\pi}{2},0)\,,\quad
(r(t),\theta(t),\dfrac{3}{2}\pi,0)\,,
\end{equation}
where $r(t)$ and $\theta(t)$ satisfy the equation
\begin{equation}\label{eqn:2mr12hrs=c1}
2mr+\dfrac{1}{2h}r^{2}\sin^{2}\theta=c_{1}\,.
\end{equation}
Here,
in the case when $m=h=1$,
Figure \ref{fig:Taub_rtheta2} shows
the solution curves of \eqref{eqn:2mr12hrs=c1} in the $(r,\theta)$-plane
for $c_{1}=1,\dotsc,10$.

\paragraph{Case 2: $c_{2}$ is a pure imaginary number}
We set $c_{2}=ic$ for some real constant $c$.
Then, we get $u(t)=ict$ by \eqref{eqn:TaubNUT_SLS_SO32nd'},
so that the fourth equation of \eqref{eqn:zu_rtpp}
yields $\psi=-(c/2m)t$.
The second equation in \eqref{eqn:TaubNUT_SLS_SO3} also becomes
\begin{equation}
\mathrm{Im}\,z=0\,,
\end{equation}
by \eqref{eqn:TaubNUT_zu_rtpp}, we get
\begin{equation}
r\sin\theta\sin\phi=0\,.
\end{equation}
A similar argument in Case 1 shows that
$L=H\cdot \ell$
gives special Lagrangian submanifolds
for the solution curves,
\begin{equation}
\ell(t)=(c_{1}/2m,0,\phi(t),-(c/2m)t)\,,\quad
(c_{1}/2m,\pi,\phi(t),-(c/2m)t)\,,
\end{equation}
or
\begin{equation}
\ell(t)=(r(t),\theta(t),0,-(c/2m)t)\,,\quad
(r(t),\theta(t),\pi,-(c/2m)t)\,,
\end{equation}
where $r(t)$ and $\theta(t)$ satisfy \eqref{eqn:2mr12hrs=c1}.
We can find the solution curves of \eqref{eqn:2mr12hrs=c1} in the $(r,\theta)$-plane
for each $c_1$.

\section{Construction of special Lagrangian submanifolds in the Atiyah-Hitchin manifold}\label{sec:AH_SLag}

In this section, we construct special Lagrangian submanifolds in Atiyah-Hitchin manifold according to Subsections \ref{sec:SLag_const} and \ref{sec:moment_SLag_const}.

\subsection{Deriving the conditions for being special Lagrangian}

Let $M$ be the Atiyah-Hitchin manifold.
We denote by $\omega$ its K\"ahler form on $M$.
In terms of the holomorphic coordinates $(U, Z)$ 
 defined in \eqref{eqn:uz_UZ}, the holomorphic volume form 
$\Omega$ is expressed as
$\Omega=du\wedge dz=dU\wedge dZ$.
Then, we have verified that
$\omega$ and $\Omega$ satisfy
the Calabi-Yau condition \eqref{eqn:CY}
in Subsection \ref{sec:AH_section}.

We consider the vector field
\begin{equation}
X=-2i\left(
Z\dfrac{\partial}{\partial Z}-\bar{Z}\dfrac{\partial}{\partial \bar{Z}}
\right)\,,
\end{equation}
which generates one parameter transformation group $H\cong SO(2)$ and acts 
 Hamiltonianly on the Atiyah-Hitchin manifold.
As shown in the case of the Taub-NUT manifold,
the corresponding moment map $\mu$
is given (up to constant)
as follows:
\begin{equation}\label{eqn:moment_AH}
\mu=-4\eta_{1}-2(x_{+}+x_{-})\omega_{1}\,.
\end{equation}
Indeed,
it is sufficient to show
$\iota_{X}\omega_{AH}=d\mu$
by verifying that
the components
of the both sides are the same.
A direct calculation shows
\begin{equation}
\iota_{X}\omega
=
K_{U\bar{Z}}\bar{Z}dU
+K_{Z\bar{U}}Zd\bar{U}
+K_{Z\bar{Z}}\bar{Z}dZ
+K_{Z\bar{Z}}Zd\bar{Z}\,.
\end{equation}
On the other hand,
since \eqref{eqn:ABI_eq123} yields
\begin{equation}
d\omega_{1}=0\,,
\end{equation}
we get
\begin{equation}
d\mu=-4d\eta_{1}-2(dx_{+}+dx_{-})\omega_{1}\,.
\end{equation}
We can read off the components
$d\eta_1$ and $dx_{\pm}$
by using \eqref{eqn:deta_UZ}
and \eqref{eqn:dxpm_UZ}, respectively.
\begin{align}
(\text{The $dU$-}\text{component of $d\mu$})
&=-\dfrac{A_{-}-A_{+}}{2(A_{-}B_{+}-A_{+}B_{-})}-\dfrac{-B_{+}+B_{-}}{A_{-}B_{+}-A_{+}B_{-}}\omega_{1} \notag \\
&=K_{U\bar{Z}}\bar{Z}
\end{align}
Here, we have used \eqref{eqn:ABI_eq144'} in the second equality.
By using \eqref{eqn:ABI_eq145'}
and \eqref{eqn:ABI_eq143'},
it can be verified that
\begin{align}
&(\text{The $d\bar{U}$-component of $d\mu$}) = K_{Z\bar{U}}Z\,,\\
&(\text{The $dZ$-component of $d\mu$}) = K_{Z\bar{Z}}\bar{Z}\,,\\
&(\text{The $d\bar{Z}$-component of $d\mu$}) = K_{Z\bar{Z}}Z\,,
\end{align}
so that $\iota_{X}\omega_{AH}=d\mu$ holds.

Following Subsection \ref{sec:moment_SLag_const},
we give special Lagrangian submanifolds with cohomogeneity-one symmetry $H$ in the Atiyah-Hitchin manifold $M$.
We first consider a curve $\ell$ in the Atiyah-Hitchin manifold $M$,
\begin{equation}
\ell:I \to M;~t\mapsto \ell(t)=(U(t),\bar{U}(t),Z(t),\bar{Z}(t))\,.
\end{equation}
We set $L=H\cdot \ell$ and define $v_{1},v_{2}\in T_{\ell(t)}L$ as follows:
\begin{align}
v_{1} &= X_{\ell(t)}=-2i\left(Z\dfrac{\partial}{\partial Z}-\bar{Z}\dfrac{\partial}{\partial \bar{Z}}\right)\,,\\
v_{2} &= \dot{\ell}(t)=
\dot{U}\dfrac{\partial}{\partial U}
+\dot{\bar{U}}\dfrac{\partial}{\partial \bar{U}}
+\dot{Z}\dfrac{\partial}{\partial Z}
+\dot{\dot{Z}}\dfrac{\partial}{\partial \dot{\bar{Z}}}\,.
\end{align}
In order to guarantee that $\ell$ intersects transversally the $H$-orbit $H\cdot \ell(t)$
at $\ell(t)$ for each $t$, $\{v_{1},v_{2}\}$ must satisfy
\begin{equation}
|Z(t)| \neq \text{constant}\,.
\end{equation}
From  \eqref{eqn:moment_AH},
for any constant value $c\in\mathbb{R}\cong Z(\mathfrak{so}(2)^{*})$,
if
\begin{equation}\label{eqn:HSO(2)AH_Lag}
-4\eta_{1}-2(x_{+}+x_{-})\omega_{1} = c\,,
\end{equation}
then $L$ becomes a Lagrangian submanifold in $M$.

Next, we rewrite the condition \eqref{eqn:condition_SLag_curve}
for $L$ to be a special Lagrangian submanifold in $M$.
We have
\begin{equation}
\Omega(v_1, v_2)=2iZ(t)\dot{U}(t)\,.
\end{equation}
Hence \eqref{eqn:condition_SLag_curve} with $\theta=0$
is equivalent to
\begin{equation}\label{eqn:AH_SLag_condtion}
\mathrm{Re}(Z(t)\dot{U}(t))=0\,.
\end{equation}
From the above argument we have obtained the
conditions for $\ell$ such that
$L=H\cdot \ell$ is a special Lagrangian submanifold
in the Atiyah-Hitchin manifold $M$.
Here, under the variable transformation \eqref{eqn:uz_UZ},
\eqref{eqn:AH_SLag_condtion} is rewritten as:
\begin{equation}
\mathrm{Re}(2\dot{u}(t)z(t)+u(t)\dot{z}(t))=0\,.
\end{equation}

\subsection{Analysis of the conditions for being special Lagrangian}
In order for the Lagrangian submanifold $L=H\cdot\ell$ to be a special Lagrange submanifold in the Atiyah-Hitchin manifold $M$,
the curve $\ell=\ell(t)$ must satisfy the following conditions: $|Z|\neq\text{constant}$ and
\begin{equation}\label{eqn:AH_SO(3)_SLAG}
-4\eta_1-2(x_{+}+x_{-})\omega_{1}=\text{constant}\,,\quad
\mathrm{Re}(Z\dot{U})=0\,.
\end{equation}
It is difficult to solve this equation
for the curve $\ell$ in general.
For simplicity, we give some special solutions
to \eqref{eqn:AH_SO(3)_SLAG}.
We first express \eqref{eqn:AH_SO(3)_SLAG} in terms of the spherical coordinates.
Here, the representations of $z,v$ and $x$ by the spherical coordinates $(k,\theta,\phi,\psi)$
 are given by \cite{IR}
\begin{equation}\label{eqn:sc_IR}
\begin{cases}
z&=2e^{2i\phi}(\cos2\psi(1+\cos^{2}\theta)+2i\sin2\psi\cos\theta
+(2k^{2}-1)\sin^{2}\theta)K^{2}(k)\,,\\
v&=8e^{i\phi}\sin\theta(\sin2\psi
-i\cos2\psi\cos\theta+i(2k^{2}-1)\cos\theta)K^{2}(k)\,,\\
x&=4(-3\cos2\psi\sin^{2}\theta+(2k^{2}-1)(1-3\cos^{2}\theta))
K^{2}(k)\,,
\end{cases}
\end{equation}
where the ranges for $\theta, \phi, \psi$ are the same ones in (\ref{eqn:zu_rtpp}).
By \eqref{eqn:dfdx=homega1} and
the first equation of \eqref{eqn:omega_1/rhoKk},
we get
\begin{equation}
\dfrac{\partial F}{\partial x}
=-\dfrac{1}{h}+\dfrac{4}{\sqrt{\rho}}K(k)\,,
\end{equation}
from which the second equation of \eqref{eqn:dFdv=v_dFdx=0}
yields
\begin{equation}
\rho=16h^{2}K^{2}(k)\,.
\end{equation}
Then, $\eta_1$ is expressed as
\begin{align}
-4\eta_{1}
&=\dfrac{4}{\sqrt{\rho}}\left\{
e_{1}K(k)-\rho E(k)
\right\}\notag \\
&=\dfrac{1}{hK(k)}\left\{
-\dfrac{16h^{2}}{3}(k^{2}-2)K^{3}(k)-16h^{2}K^{2}(k)E(k)
\right\}\notag \\
&=-16hK(k)\left\{
\dfrac{1}{3}(k^{2}-2)K(k)+E(k)
\right\}\,.
\end{align}
Here, we have used \cite[(A.14)]{Arai:2022xyc} in the first equality.
From \eqref{eqn:xpm_escpt}, we have
\begin{multline}
x_{\pm}
= \dfrac{K^{2}(k)}{3}\bigg[-12\cos2\psi\sin^{2}\theta+4(2k^{2}-1)(1-3\cos^{2}\theta)\\
\pm12\left\{(\cos2\psi(1+\cos^{2}\theta)+(2k^{2}-1)\sin^{2}\theta)^{2}+4\sin^{2}2\psi\cos^{2}\theta\right\}^{1/2}\bigg]\,,
\end{multline}
so that we find
\begin{equation}
-2(x_{+}+x_{-})\omega_{1}
=\dfrac{4}{3h}K^{2}(k)\left\{
3\cos2\psi\sin^{2}\theta-(2k^{2}-1)(1-3\cos^{2}\theta)
\right\}\,.
\end{equation}
Therefore, the first equation of \eqref{eqn:AH_SO(3)_SLAG} is expressed as
\begin{multline}\label{eqn:AH_SO(3)_SLAG1st'}
-16hK(k)\left\{
\dfrac{1}{3}(k^{2}-2)K(k)+E(k)
\right\}\\+
\dfrac{4}{3h}K^{2}(k)\left\{
3\cos2\psi\sin^{2}\theta-(2k^{2}-1)(1-3\cos^{2}\theta)
\right\}=c_{1}\,,
\end{multline}
where $c_{1}$ is a constant.

In what follows, we restrict our considerations to
the case where $\dot{U}(t)$ is a real constant:
\begin{equation}\label{eqn:AH_SLS_Udotassume}
\dot{U}(t)=c_{2}\qquad
\text{($c_{2}$: constant)}\,.
\end{equation}
This can be expressed by using the holomorphic coordinate $(u,z)$ as
\begin{equation}
u\sqrt{z}=c_{2}t+c_{3}\,. \label{eqn:secAH}
\end{equation}
Then, the second equation of \eqref{eqn:AH_SO(3)_SLAG} is
rewritten as:
\begin{equation}\label{eqn:rez=0}
\mathrm{Re}\sqrt{z}=0\,.
\end{equation}
By using the spherical coordinates as in \eqref{eqn:sc_IR}, 
\eqref{eqn:rez=0} is also rewritten as
\begin{multline}\label{eqn:AH_ImOmega_spherecord}
e^{i\phi}K(k)\left\{
\cos2\psi(1+\cos^{2}\theta)+(2k^{2}-1)\sin^{2}\theta+2i\sin2\psi\cos\theta
\right\}^{1/2}\\
+e^{-i\phi}K(k)\left\{
\cos2\psi(1+\cos^{2}\theta)+(2k^{2}-1)\sin^{2}\theta-2i\sin2\psi\cos\theta
\right\}^{1/2}=0\,.
\end{multline}

We are now ready to solve the conditions \eqref{eqn:AH_SO(3)_SLAG} written by the spherical coordinates,
 namely, \eqref{eqn:AH_SO(3)_SLAG1st'} and \eqref{eqn:AH_ImOmega_spherecord}.
We derive curve solutions satisfying these conditions.
We consider two cases. 
In the first case, we eliminate $\psi$ by substituting \eqref{eqn:AH_SO(3)_SLAG1st'} 
 into \eqref{eqn:AH_ImOmega_spherecord} and so two conditions reduce to one condition.
We show the solution curves of that condition in $(\theta,\phi)$-plane with fixing $k$.
In the second case, we plot the solution curves for the same condition in $(\theta, k)$-plane with fixing $\phi$.
$\psi$ can be eliminated by solving \eqref{eqn:AH_SO(3)_SLAG1st'} with respect to $\cos2\psi$:
\begin{multline}\label{eqn:AH_elmi_psi}
\cos2\psi
=\dfrac{1}{3\sin^{2}\theta}\bigg\{
(2k^{2}-1)(1-3\cos^{2}\theta)\\
+\dfrac{3h}{4K^{2}(k)}\left(
c_{1}+16hK(k)\left(
\dfrac{1}{3}(k^{2}-2)K(k)+E(k)
\right)
\right)
\bigg\}\,.
\end{multline}

\paragraph{Case 1: Solution curves in $(\theta,\phi)$-plane}

\begin{figure}[H]
\centering
\begin{tabular}{ccc}
\includegraphics[keepaspectratio, scale=0.3]{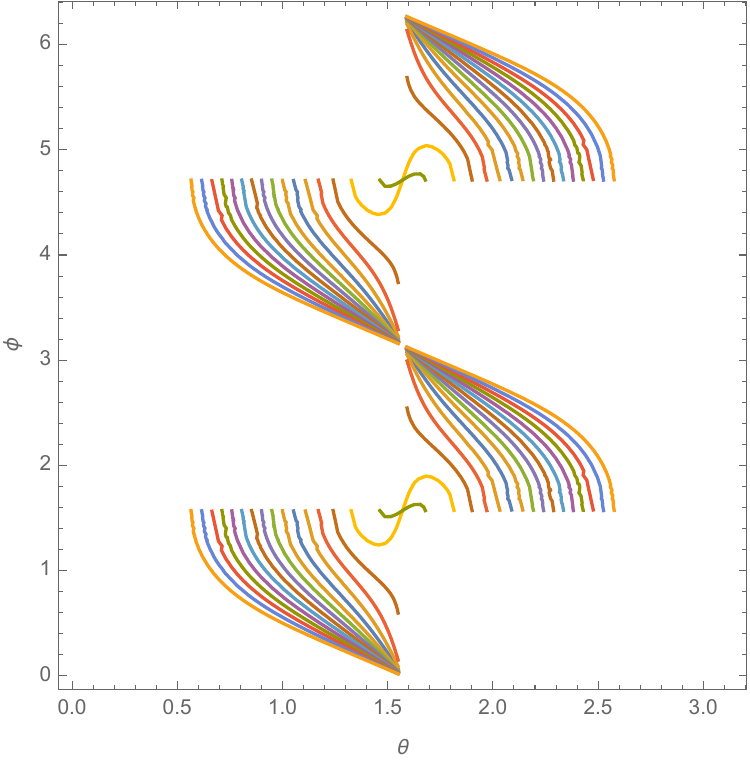}&
\includegraphics[keepaspectratio, scale=0.3]{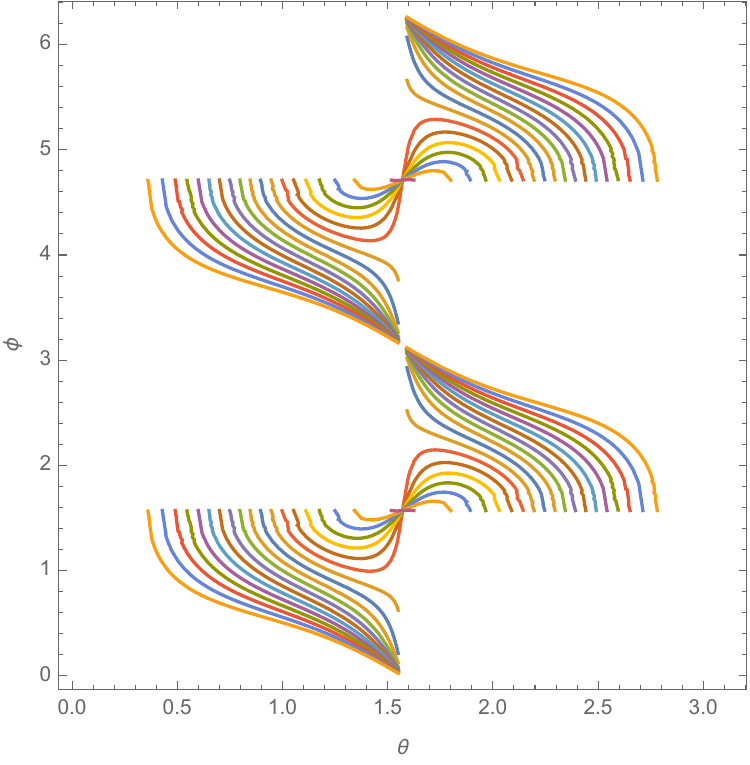}&
\includegraphics[keepaspectratio, scale=0.3]{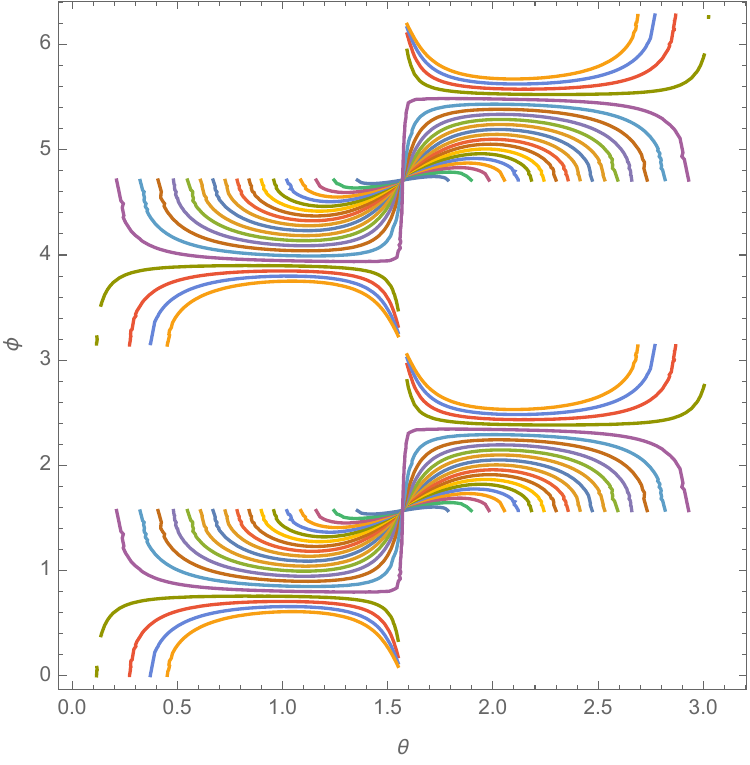}\\
$k=0.3$ & $k=0.5$ & $k=0.7$
\end{tabular}
\caption{Solution curves for 
\eqref{eqn:AH_ImOmega_spherecord}
with eliminating $\psi$ in the $(\theta,\phi)$-plane.}\label{fig:AH_thetaphi}
\end{figure}

We can eliminate $\psi$ from 
\eqref{eqn:AH_ImOmega_spherecord} by using \eqref{eqn:AH_elmi_psi}.
Figure \ref{fig:AH_thetaphi} shows solution curves for
the obtained equation with respect to $(\theta,\phi)$
with $h=1$, $k=0.3$, $0.5$, $0.7$ and $c_{1}=0$, $\pm1$, $\dotsc$, $\pm10$.

\begin{figure}[H]
\centering
\begin{tabular}{ccc}
\includegraphics[keepaspectratio, scale=0.3]{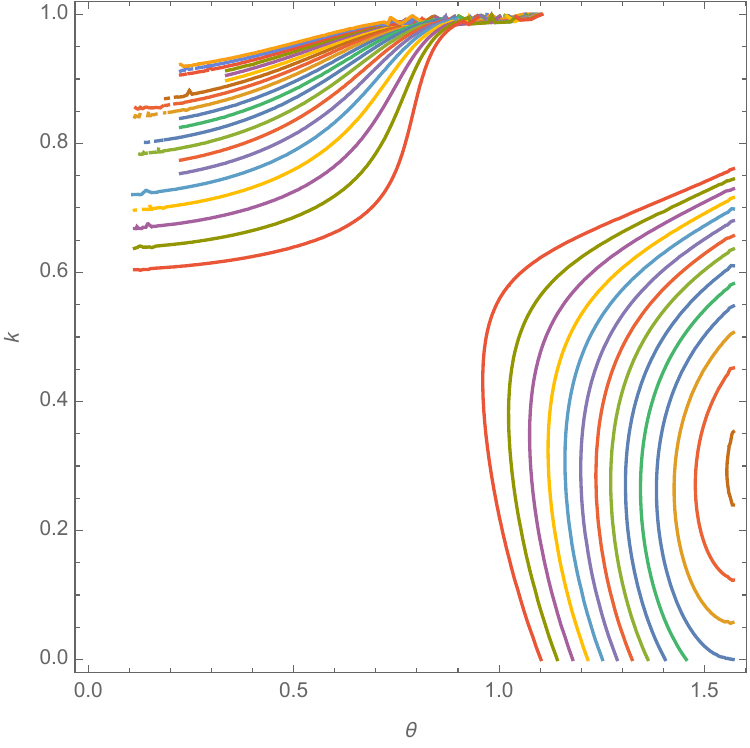}&
\includegraphics[keepaspectratio, scale=0.3]{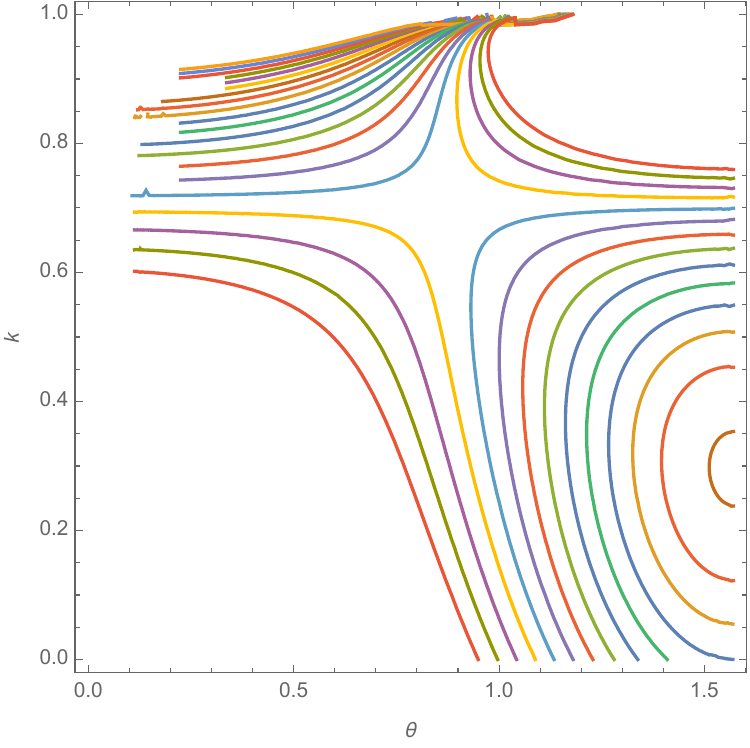}&
\includegraphics[keepaspectratio, scale=0.3]{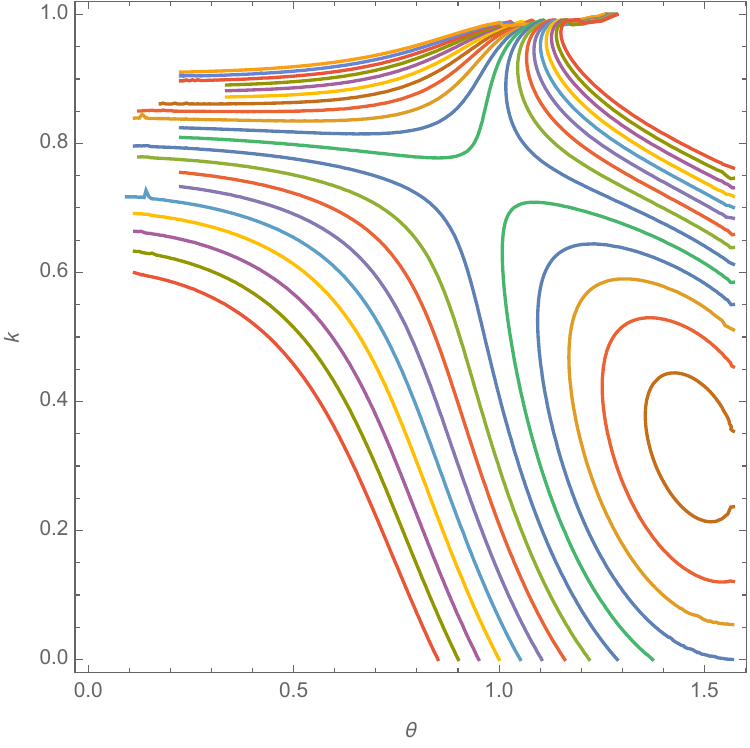}\\
$\phi=\dfrac{\pi}{6}$ &
$\phi=\dfrac{\pi}{4}$ &
$\phi=\dfrac{\pi}{3}$
\end{tabular}
\caption{Solution curves for 
\eqref{eqn:AH_ImOmega_spherecord}
with eliminating $\psi$ in the $(\theta,k)$-plane.}\label{fig:AH_kpsi}
\end{figure}

\paragraph{Case 2: Solution curves in $(\theta,k)$}
We make use the same equation in Case 1.
Then, Figure \ref{fig:AH_kpsi}
shows solution curves for
the equation with respect to $(\theta,k)$
with $\phi=\pi/6$, $\pi/4$, $\pi/3$, 
$c_{1}=0$, $\pm1$, $\dotsc$, $\pm10$
and $h=1$.

%
%
\section{Summary and conclusion}\label{sec:summary}

Calibrated submanifolds have been actively studied in both mathematics and physics since their introduction
 by Harvey and Lawson \cite{HL}.
As one class of calibrated submanifolds, this paper have discussed  
 special Lagrangian submanifolds.
Especially, we have focused on the construction of special Lagrangian submanifolds in 
 the Taub-NUT manifold and the Atiyah-Hitchin manifold.
In order to construct them, we have combined the generalized Legendre transform approach and the moment map technique.
We have seen that the generalized Legendre transform approach naturally gives
 the description of their Calabi-Yau structures in terms of holomorphic coordinates,
 which are useful to employ the moment map technique
and to describe the conditions for submanifolds to be special Lagrangian submanifolds.
We have reviewed the necessary tools for the moment map technique such as the K\"ahler metrics of 
 the Taub-NUT manifold and the Atiyah-Hitchin manifold which are obtained in the generalized Legendre transform approach.
While they are derived from the $F$-functions given in \cite{IR} and \cite{Hitchin:1986ea}, respectively,
 their detailed derivation was especially necessary for the Atiyah-Hitchin manifold.
That has been worked out in our previous paper \cite{Arai:2022xyc} and we have reviewed this in Subsection 
 \ref{sec:AH_section}.
In this paper, we have explained the moment map technique including the underlying concepts like Hamiltonian actions
 and cohomogeneity-one actions.
By the moment map technique,  we have derived the condition for a submanifold $L$ which is obtained as a one-parameter family of the
 orbits for an Hamiltonian action to be special Lagrangian.
In the derivation, we have evaluated the holomorphic volume form of such submanifolds
 which are readily obtained in the generalized Legendre transform approach.
By applying the moment map technique to the Taub-NUT manifold and the Atiyah-Hitchin manifold,
 we have derived conditions for $L$ to be special Lagrangian which are described by ODEs 
 with respect to the
 one parameter.
These ODEs are difficult to solve by quadrature methods in general.
We have solved the ODEs numerically and have plotted solutions curves,
 which correspond to the cohomogeneity-one special Lagrangian submanifolds in the above manifolds.

%
%
\subsubsection*{}
\noindent {\bf Acknowledgements} \\
\noindent This work is supported in part by JSPS Grant-in-Aid for Scientific  
Research KAKENHI Grant No. JP21K03565 (M. A.).

%
%


\begin{thebibliography}{99}
%
\bibitem{HL}
F.~R.~Harvey and H. B.~Lawson, 
 ``Calibrated geometries," 
 Acta Math. {\bf 148} (1982), 47-157.
%
\bibitem{HL2}
F.~R.~Harvey,
{\it Spinors and calibrations},
Academic Press, 1990.
%
\bibitem{HL3}
F.~R.~Harvey and M.~L.~Michelsohn,
{\it Spin geometry},
Princeton University Press, 1989.
%
\bibitem{Figueroa-OFarrill:1998kci}
J.~M.~Figueroa-O'Farrill,
 ``Intersecting brane geometries,''
J. Geom. Phys. \textbf{35} (2000), 99-125
[arXiv:hep-th/9806040 [hep-th]].
%
\bibitem{SYZ}
A.~Strominger, S.~T.~Yau and E.~Zaslow, 
``Mirror symmetry is T-duality," 
Nucl. Phys. B \textbf{479} (1996) 243-259.
%
\bibitem{mirror2}
D.~R.~Morrison,
``The Geometry underlying mirror symmetry,''
[arXiv:alg-geom/9608006 [math.AG]].
%
\bibitem{ENOOST}
M.~Eto, Y.~Isozumi, M.~Nitta, K.~Ohashi, K.~Ohta, N.~Sakai and Y.~Tachikawa,
``Global structure of moduli space for BPS walls,''
Phys. Rev. D \textbf{71} (2005), 105009
[arXiv:hep-th/0503033 [hep-th]].
%
\bibitem{joyce}
D.~Joyce,
``Special Lagrangian $m$-folds in $\mathbb C^m$ with symmetries,"
 Duke Math. J. {\bf 115} (2002) 1-51.
%
\bibitem{stenzel1}
M.~B.~Stenzel, 
``K\"ahler Structures on the cotangent bundles of real analytic riemannian manifolds,"
 Ph.D. Thesis, Massachusetts Institute of Technology, 1990.
%
\bibitem{stenzel2}
M.~B.~Stenzel,
``Ricci-flat metrics on the complexification of a compact rank one symmetric space,"
Manuscripta Math. \textbf{80} (1993), no. 2, 151-163.
%
\bibitem{anciaux}
H.~Anciaux, 
``Special Lagrangian submanifolds in the complex sphere,"
Ann. Fac. Sci. Toulouse Math. (6), \textbf{16}, (2007), no. 2, 215-227.
%
\bibitem{IM}
M.~Ionel and M.~Min-Oo, 
``Cohomogeneity one special Lagrangian 3-folds in the deformed and the resolved conifolds,"
Illinois J. Math. \textbf{52} (2008), no. 3, 839-865.
%
\bibitem{HS}
K.~Hashimoto and T.~Sakai, 
``Cohomogeneity one special Lagrangian submanifolds in the cotangent bundle of the sphere," 
Tohoku Math. J. (2) \textbf{64} (2012), no. 1, 141-169.
%
\bibitem{HM}
K.~Hashimoto and K.~Mashimo,
``Special Lagrangian submanifolds invariant under the isotropy action of symmetric spaces of rank two,"
J.~Math Soc.~Japan \textbf{68} (2016), 839-862.
%
\bibitem{AB1}
M.~Arai and K.~Baba,
``Special Lagrangian Submanifolds and Cohomogeneity One Actions on the Complex Projective Space,"
 Tokyo J. Math. \textbf{42} (2019), 255-284.
%
\bibitem{Koike}
N.~Koike,
``Calabi-Yau structures and special Lagrangian submanifolds of complexified symmetric spaces,"
Illinois Journal of Mathematics \textbf{63} (2019), 575-600.
%
\bibitem{Eguchi:1978gw}
T.~Eguchi and A.~J.~Hanson,
``Selfdual Solutions to Euclidean Gravity,''
Annals Phys. \textbf{120} (1979), 82.
%
\bibitem{Gibbons:1979xm}
G.~W.~Gibbons and S.~W.~Hawking,
``Classification of Gravitational Instanton Symmetries,''
Commun. Math. Phys. \textbf{66} (1979), 291-310.
%
\bibitem{Taub:1950ez}
A.~H.~Taub,
``Empty space-times admitting a three parameter group of motions,''
Annals Math. \textbf{53} (1951), 472-490.
%
\bibitem{Newman:1963yy}
E.~Newman, L.~Tamburino and T.~Unti,
``Empty space generalization of the Schwarzschild metric,''
J. Math. Phys. \textbf{4} (1963), 915.
%
\bibitem{Atiyah:1988jp}
M.~F.~Atiyah and N.~J.~Hitchin,
``The geometry and dynamics of magnetic monopoles,''
Princeton University Press, Princeton, NJ, 1988.
%
\bibitem{Noda}
T.~Noda,
 ``A special Lagrangian fibration in the Taub-NUT space,"
 J. Math. Soc. Japan \textbf{60}(3), (2008) 653-663.
%
\bibitem{Lindstrom:1983rt}
U.~Lindstrom and M.~Rocek,
``Scalar Tensor Duality and N=1, N=2 Nonlinear Sigma Models,''
Nucl. Phys. B \textbf{222} (1983), 285-308.
%
\bibitem{Hitchin:1986ea}
N.~J.~Hitchin, A.~Karlhede, U.~Lindstrom and M.~Rocek,
``Hyperkahler Metrics and Supersymmetry,''
Commun. Math. Phys. \textbf{108} (1987), 535.
%
\bibitem{Karlhede:1986mg}
A.~Karlhede, U.~Lindstrom and M.~Rocek,
``Hyperkahler Manifolds and Nonlinear Supermultiplets,''
Commun. Math. Phys. \textbf{108} (1987), 529.
%
\bibitem{Lindstrom:1987ks}
U.~Lindstrom and M.~Rocek,
``New Hyperkahler Metrics and New Supermultiplets,''
Commun. Math. Phys. \textbf{115} (1988), 21.
%
\bibitem{IR}
  I.~T.~Ivanov and M.~Rocek,
  ``Supersymmetric sigma models, twistors, and the Atiyah-Hitchin metric,''
  Commun.\ Math.\ Phys.\  {\bf 182} (1996) 291
  [hep-th/9512075].
%
\bibitem{Ionas1}
 R.~A.~Ionas,
``Elliptic constructions of hyperkaehler metrics. I. The Atiyah-Hitchin manifold,''
[arXiv:0712.3598 [math.DG]].
%
\bibitem{Ionas2}
  R.~A.~Ionas,
  ``Elliptic constructions of hyperkahler metrics. III. Gravitons and Poncelet polygons,''
  arXiv:0712.3601 [math.DG].
%
\bibitem{Arai:2022xyc}
M.~Arai, K.~Baba and R.~A.~Ionas,
``Revisiting Atiyah\textendash{}Hitchin manifold in the generalized Legendre transform,''
PTEP \textbf{2023} (2023) no.6, 063A03
[arXiv:2206.02420 [hep-th]].
%
\bibitem{Bielawski}
R.~Bielawski,
``Line bundles on spectral curves and the generalised Legendre transform
construction of hyperk\"ahler metrics,''
J.~Geom.~Phys.
\textbf{59} (2009), 374--390.
%
\bibitem{Eguchi:1980jx}
T.~Eguchi, P.~B.~Gilkey and A.~J.~Hanson,
``Gravitation, Gauge Theories and Differential Geometry,''
Phys. Rept. \textbf{66} (1980), 213.
%
\end{thebibliography}
\end{document}